\documentclass[final]{elsarticle}

\makeatletter
\def\ps@pprintTitle{%
 \let\@oddhead\@empty
 \let\@evenhead\@empty
 \def\@oddfoot{}%
 \let\@evenfoot\@oddfoot}
\makeatother
\usepackage{hyperref}
\usepackage[dvipsnames]{xcolor}
\usepackage{amsmath}
\usepackage{amsbsy}
\usepackage{amsthm}
\usepackage{amssymb}
\usepackage[tmargin=1in,bmargin=1in]{geometry}
\usepackage{subfigure}
\usepackage{verbatim}
\usepackage{graphicx}
\usepackage{latexsym}
\usepackage{pdfsync}
\usepackage{multirow}
\usepackage{rotating}
\usepackage{color}
\usepackage{caption}
\usepackage{url}
 \global\bibsep=0pt
\usepackage[normalem]{ulem}
\usepackage{framed,multirow}

\usepackage{url}
\usepackage{xcolor} 
\usepackage[normalem]{ulem}
\usepackage[final]{pdfpages}

\usepackage{algorithm}
\usepackage{float}
\newfloat{algorithm}{t}{lop}
\floatname{algorithm}{Scheme}

\newcommand{\sqrtfacdVdt}{{\sqrt{\frac{2k_\mathrm{B}T\nu\Delta t}{\rho\Delta V}}}}

\usepackage{bm}
\hypersetup{
    colorlinks=true,      		
    linkcolor=blue,       		
    citecolor=blue,       		
    filecolor=black,      		
    urlcolor=red,       		  
    bookmarks=false,
    pdffitwindow=true,
    pdfpagelayout=SinglePage
}

\newcommand{\ie}{{\it i.e.\ }}
\newcommand{\eg}{{\it e.g.\ }}

\newcommand{\vertiii}[1]{{\left\vert\kern-0.25ex\left\vert\kern-0.25ex\left\vert #1 
    \right\vert\kern-0.25ex\right\vert\kern-0.25ex\right\vert}}

\definecolor{newcolor}{rgb}{.8,.349,.1}

\theoremstyle{plain}

\journal{Journal of Computational Physics}
\begin{document}

\begin{frontmatter}
\title{Projection Method for the Fluctuating Hydrodynamics Equations }
\journal{Journal of Computational Physics}
\address[X]{{Département de Mathématiques Appliquées,} Ecole polytechnique, 91128 Palaiseau, France.}
\address[UCMERCED]{Department of Applied Mathematics, University of California, Merced, California 95343, USA.}
\author[X]{Marc Mancini}
\author[UCMERCED]{Maxime Theillard} 
\author[UCMERCED]{Changho Kim}

\cortext[cor]{Corresponding author: ckim103@ucmerced.edu}

\begin{abstract}

Computational fluctuating hydrodynamics aims at understanding the impact of thermal fluctuations
on fluid motions at small scales through numerical exploration. 
These fluctuations are modeled as stochastic flux terms and incorporated into the classical Navier--Stokes
equations, which need to be solved numerically.
In this paper, we present a novel projection-based method for solving the incompressible fluctuating
hydrodynamics (FHD) equations. 
By analyzing the equilibrium structure factor spectrum of the velocity field for the linearized FHD equations, we investigate 
how the inherent splitting errors affect the numerical solution of the stochastic partial differential
equations in the presence of non-periodic boundary conditions, and how iterative corrections can reduce these errors.
Our computational examples demonstrate both the capability of our approach to   reproduce correctly
stochastic properties of fluids at small scales as well as its potential use in the simulations of
multi-physics problems.

\end{abstract}


\end{frontmatter}

\section{\label{sec_intro} Introduction}

This paper presents a projection-based method for solving the incompressible fluctuating hydrodynamics equations, with the dual intent to provide an alternative simulation strategy and illustrate how traditional computational fluid dynamics techniques can be adapted to incorporate thermal fluctuations.

\subsection{Fluctuating Hydrodynamics Approach}

Owing to the perpetual miniaturization of engineered systems, 
the smallest scale of interest for engineers and scientists has been continuously shrinking, 
leading to the emergence of nanometric devices and a growing curiosity 
into the fundamental properties of nanofluidic systems~\cite{nanofluidicOnTheRise20201}. 
This trend has brought various exciting perspectives and novel ideas to the scientific community, including bioengineered kinesin-microtubule systems~\cite{lopes_membrane_2019}, 
nanofabricated devices for biomolecule applications~\cite{B917759K},
nano heat transport technologies~\cite{nanoheatreview, Wang2009}, 
brain-on-a-chip platforms~\cite{10.3389/fbioe.2019.00100}, 
and micro-sensing devices~\cite{microsensing}, to name but a few.
These systems display exotic behaviors and acquire unique features because their characteristic length scales are small, for example, comparable to the Debye length, the size of biomolecules, or even the slip length~\cite{Abgrall2008}.

In particular, since the characteristic length and time scales of the systems investigated are no longer widely separated from those of the underlying molecular systems, 
the validity of any deterministic continuum models is questionable and should be investigated. Surprisingly, experimental~\cite{doi:10.1063/1.449693, doi:10.1063/1.465059, doi:10.1126/science.1074481, PhysRevB.75.115415, PhysRevLett.96.086105, PhysRevLett.91.166104}, 
theoretical~\cite{B909366B}, 
and numerical~\cite{molhydro, PhysRevLett.102.184502, B616490K, PhysRevLett.94.026101} 
studies have found that the Navier--Stokes equations remain largely valid down to a few nanometers. 
Specifically, the classical hydrodynamics equations can predict the \textit{average} nanofluidics behavior near the equilibrium or in simple steady states even at the nanometer scale. 
However, thermal fluctuations appearing in the instantaneous continuum fields are no longer negligible as the system size becomes smaller.
When these fluctuations interact with nonlinearity in the system, the entire dynamics of the fluid system
may not be correctly captured by the deterministic continuum description.
For example, in the giant fluctuation experiment in space~\cite{GFexp,GFexp2}, 
concentration fluctuations have been observed to grow to the macroscale
(\ie sizes ranging up to millimeters and relaxation times as large as hundreds of seconds)
under the presence of the concentration gradient due to the coupling with random advection.
Therefore, for a complete and accurate representation of nanofluidic and sub-nanofluidic systems, 
thermal fluctuations must be accounted for.

The use of stochastic partial differential equations (SPDEs) to describe fluid dynamics
dates back to Landau and Lifshitz~\cite{LandauLifshtz1959}.
They proposed to incorporate stochastic fluxes to each dissipative process (\eg momentum diffusion) 
in the Navier--Stokes equations to correctly model the effects of thermal fluctuations.
This fluctuating hydrodynamics (FHD) approach has been successfully applied to
describe various phenomena induced by hydrodynamic fluctuations.
Until significant progress in the computational FHD approach has been made for the past two decades
(see below),
most accomplishments of the FHD theory were made by analytical methods~\cite{OrtizDeZarateSengers2006}.
While analytical approaches have provided insightful explanations,
they are mostly limited to simple nonequilibrium situations and 
moreover, they rely on several assumptions (\eg linearization and periodic boundary conditions).
We note that stochastic terms can also be added to the deterministic fluid equations in the context of
uncertainty quantification~\cite{MathelinHussainiZang2005, Wan2007} 
to represent a large variety of noisy contributions 
(\eg uncertainty on the boundary conditions~\cite{PhysRevLett.92.154501}).

As mentioned above, significant progress in the computational FHD approach has been made
for the past two decades.
Here we focus on the PDE-based (as opposed to particle-based) approaches for homogeneous fluids.
The Landau--Lifshitz Navier--Stokes equations (\ie compressible FHD equations) were numerically solved 
for the one-species~\cite{PhysRevE.75.026307,CCSE_PRE2007} and binary mixture~\cite{CCSE_ESAIM2010} cases
as well as the multi-species case~\cite{CCSE_PRE2014}, which was extended to include
stochastic reactions~\cite{CCSE_JCP2015}.
The incompressible FHD formulation~\cite{BalboaUsabiagaBellDelgadoBuscalioniDonvFaiGriffithPeskin2012}
was extended to quasi-incompressible fluids
by the low Mach number formulation for the binary mixture~\cite{CCSE_CAMCOS2014, CCSE_CAMCOS2015} 
and muiti-species cases~\cite{CCSE_PhysFluids2015}.
The quasi-incompressible case was extended to include stochastic reactions~\cite{CCSE_JChemPhys2018}
as well as charges (\ie electrolyte ions)~\cite{CCSE_PRF2016, CCSE_PRF2019}.
To construct and analyze numerical schemes for FHD equations,
various advanced deterministic PDE and computational fluid dynamics (CFD) techniques 
have been applied and extended.
Spatial discretization based on the finite-volume approach and the stochastic version of the method of
lines were introduced,
and the structure factor analysis technique was developed for the systematic construction 
of stochastic numerical methods~\cite{DonevVandenEijndenGarciaBell2010}.
For (quasi-)incompressible FHD equations, staggered spatial discretization schemes 
(\ie using a grid with staggered momenta) were
developed~\cite{BalboaUsabiagaBellDelgadoBuscalioniDonvFaiGriffithPeskin2012, VoulgarakisChu2009}.
In addition, to solve the stochastic Stokes problem, which is a saddle-point linear system, 
using the generalized minimal residual method (GMRES), an efficient variable-coefficient 
finite-volume Stokes solver was developed~\cite{CaiNonakaBellGriffithDonev2014}.
Several time integrators for FHD equations (\eg semi-implicit schemes~\cite{CCSE_CAMCOS2015})
were also constructed and analyzed~\cite{DelongGriffithVandenEijndenDonev2013}.
While it is not possible to give a complete summary of
applications and extensions of the FHD approach here,
we note that the FHD approach has been applied to
reaction-diffusion systems~\cite{Atzberger2010,ATZBERGER201357, CCSE_JChemPhys2017}
and coupled to kinetic Monte Carlo~\cite{SelmiMitchellManoranjanVoulgarakis2017} and
molecular dynamics~\cite{VoulgarakisChu2009}.
Hydrodynamic couplings between microstructures or ions were also considered
using direct numerical simulations \cite{SHARMA2004466}, the stochastic Eulerian-Lagrangian 
method~\cite{Atzberger2011,ATZBERGER201357, wang_fluctuating_2018,doi:10.1137/15M1019088}, the boundary
integral formulation~\cite{BaoRachhKeavenyGreengardDonev2018},
and the immersed boundary approach~\cite{ATZBERGER20071255,ATZBERGER2008379,CCSE_PRF2021}.

\subsection{Projection Method}

The projection method~\cite{CHORIN196712} uses the Hodge decomposition to decouple the fluid equations
and update the solution in two steps. 
First, the momentum contributions are used to advance the velocity field, 
which is then projected on the divergence-free space to enforce the incompressibility condition and
recover the pressure field. 
Owing to this decoupling, the projection method alleviates the need for constructing and 
solving a monolithic system containing the coupled hydrodynamic equations
by forming and solving two smaller systems. 
Using traditional data structures and discretization strategies~\cite{GUITTET2015}, 
one can guarantee that these systems are symmetric positive definite and 
therefore can efficiently solve them using classical iterative methods, 
which can be accelerated using parallel architectures~\cite{EGAN2021110084}.
On periodic domains, particularly in the context of FHD calculations \cite{ATZBERGER201357,ATZBERGER20071255}, the projection operator can be efficiently computed using fast Fourier transforms.
However, when physical boundary conditions such as the no-slip boundary condition are used, 
it is well known that decoupling causes errors near the boundaries~\cite{ELiu1995, ELiu2003}.
To avoid this issue, (semi-)implicit schemes for solving the incompressible FHD equations
have been developed mostly by solving the coupled system~\cite{CCSE_CAMCOS2015}
(note, however, that the projection approach is still employed 
as a preconditioner~\cite{CaiNonakaBellGriffithDonev2014}).
Nevertheless, since the computational advantage of using the projection method in FHD simulations is
expected to be significant, particularly for large-scale simulations,
we propose a projection-based method with iterative boundary corrections 
and perform a systematic numerical analysis based on the equilibrium structure factor.

Analytic structure factor analysis studies for incompressible FHD equations on staggered grids with non-periodic boundary conditions have been limited (see Appendix~B of Ref.~\cite{BalboaUsabiagaBellDelgadoBuscalioniDonvFaiGriffithPeskin2012}).
Using the static structure factor for the linearized FHD equations in equilibrium, we develop a semi-analytic approach to investigate projection-based methods.
While we present our projection method for the linearized FHD equations in equilibrium for the sake of a clear presentation and analysis, this method can be used to solve incompressible FHD equations that include the advection term and are coupled with concentrations in a nonequilibrium setting.
To demonstrate this, we present giant fluctuation simulation results.

\bigskip

The rest of the present paper is organized as follows.
We start in section~\ref{sec_incomp_stokes_eqn} by introducing the stochastic incompressible 
Stokes equation and its spatial discretization on uniform staggered grids.
In section~\ref{sec:cov_and_struct_fac}, we introduce the steady-state covariance and static structure factor that will be used to analyze our numerical schemes.
Our projection method is presented in section~\ref{sec_construct_methods}, analyzed in section~\ref{sec:sscov}, and  numerically validated
in section~\ref{sec_num_val}.
In section~\ref{sec_giant_fluct}, we employ our method to simulate giant fluctuations. We conclude in section~\ref{sec_conclusions}.


\section{\label{sec_incomp_stokes_eqn}Linearized Fluctuating Hydrodynamic Equations}

We introduce here an SPDE for the velocity field of an incompressible fluid and discuss its spatial and temporal discretizations.
We consider the equilibrium case, where the SPDE can be linearized.
The resulting stochastic incompressible Stokes equation is presented in section~\ref{subsec_conti_eqn}.
Its spatial discretization based on the finite-volume approach is described 
in section~\ref{subsec_spat_disc}.
As an example of a numerical scheme that does not use the projection method approach,  
a temporal integration scheme based on the Crank--Nicolson approximation
is given in section~\ref{subsec_temp_disc}.

\subsection{\label{subsec_conti_eqn}Continuum Equation}

The fluctuating behavior of an incompressible fluid in equilibrium can be 
modeled by the following FHD equations:
\begin{subequations}
\label{NS_vel_eq}
\begin{equation}
    \rho \displaystyle \frac{\partial \mathbf{u}}{\partial t} 
    + \rho \mathbf{u} \cdot \bm{\nabla}\mathbf{u} 
    = \mu \Delta \mathbf{u} - \bm{\nabla}\pi 
    + \sqrt{2k_\mathrm{B}T\mu} \bm{\nabla} \cdot \mathbf{\Sigma},
\end{equation}
\begin{equation}
    \bm{\nabla} \cdot \mathbf{u} = 0,
\end{equation}
\end{subequations}
where $\mathbf{u}(\mathbf{r},t)$ and $\pi(\mathbf{r},t)$ denote respectively
the velocity and pressure fields, $\rho$ and $T$ are respectively 
the mass density and temperature of the fluid and are taken as constant, 
$k_\mathrm{B}$ is the Boltzmann constant, and $\mu$ is the dynamic viscosity.
The tensor field $\sqrt{2k_\mathrm{B}T\mu} \mathbf{\Sigma}$ denotes the
stochastic momentum flux. 

When velocity fluctuations can be assumed to be small,
the self-advected term $\mathbf{u}\cdot\bm{\nabla}\mathbf{u}$,
which is of second order in $\mathbf{u}$, can be neglected (see \eg \cite{doi:10.1137/15M1019088} for in-depth analysis).
Under this assumption, using the kinematic viscosity $\nu = \mu / \rho$ and the orthogonal 
projection $\mathcal{P}$ onto the space of divergence-free velocity fields, 
we linearize equation \eqref{NS_vel_eq} as
\begin{equation}
\label{Stokes_vel_eq}
    \frac{\partial \mathbf{u}}{\partial t} 
    = \mathcal{P}\left[ \nu \Delta \mathbf{u} 
    + \sqrt{\frac{2k_\mathrm{B}T\nu}{\rho}} \bm{\nabla} 
    \cdot \boldsymbol{\mathcal{W}}\right].
\end{equation}
Here, we assume that $\boldsymbol{\mathcal{W}}$ is a spatiotemporal Gaussian white noise
tensor field with independent components having covariance
\begin{equation}
    \langle \mathcal{W}_{ij}(\mathbf{r},t) \mathcal{W}_{i'j'}(\mathbf{r}',t') \rangle 
    = \delta_{ii'} \delta_{jj'} \delta(\mathbf{r} - \mathbf{r}')\delta(t - t').
\end{equation}
In principle, the symmetrized form (\ie 
$\mathbf{\Sigma}=\frac{1}{\sqrt{2}}(\boldsymbol{\mathcal{W}}+\boldsymbol{\mathcal{W}}^T)$
should have been used.
However, what matters in the Fokker--Planck description is the covariance of 
the projected stochastic forcing 
$\mathcal{P}\left[\bm{\nabla} \cdot \mathbf{\Sigma}\right]$ (see \eqref{FDB_PD_L} below)
and the use of the symetrized form is not necessary for incompressible flow 
with constant viscosity~\cite{BalboaUsabiagaBellDelgadoBuscalioniDonvFaiGriffithPeskin2012,
DelongGriffithVandenEijndenDonev2013}.
We also assume that the initial velocity $\mathbf{u}(\mathbf{r},0)$ is divergence-free,
\ie $\mathcal{P}\mathbf{u}(0) = \mathbf{u}(0)$.
This assumption implies that $\mathbf{u}(\mathbf{r},t)$ remains to be divergence-free 
for all $t\ge 0$.

In this paper, we consider the periodic boundary condition and 
the no-slip boundary condition (\ie $\mathbf{u}=0$ on the boundary).
For both boundary conditions, the divergence operator $\mathcal{D}$ and 
gradient operator $\mathcal{G}$ satisfy $\mathcal{D}^* = -\mathcal{G}$,
where star denotes an adjoint of a matrix or linear operator.
Hence, the vector Laplacian operator
$\mathcal{L} = \mathcal{D}\mathcal{G}$
is self-adjoint, \ie $\mathcal{L}^* = \mathcal{L}$, and the projection operator
$\mathcal{P}=\mathcal{I}-\mathcal{G}(\mathcal{DG})^{-1}\mathcal{D}$
with $\mathcal{I}$ being the identity operator is indeed an orthogonal projection, 
\ie $\mathcal{P}^2 = \mathcal{P}$ and $\mathcal{P}^* = \mathcal{P}$.
In addition, using
$\langle \boldsymbol{\mathcal{W}}(t) \boldsymbol{\mathcal{W}}^*(t') \rangle 
= \delta(t-t')\mathcal{I}$ 
and $\mathcal{L} = -\mathcal{D}\mathcal{D}^*$,
the covariance of the projected stochastic forcing is expressed
as~\cite{BalboaUsabiagaBellDelgadoBuscalioniDonvFaiGriffithPeskin2012}
\begin{equation}
\label{FDB_PD_L}
    \langle (\mathcal{P}\mathcal{D}\boldsymbol{\mathcal{W}}(t))
    (\mathcal{P}\mathcal{D}\boldsymbol{\mathcal{W}}(t'))^* \rangle
    = -\delta(t-t') \mathcal{P} \mathcal{L} \mathcal{P}.
\end{equation}
Therefore, it can be seen in \eqref{Stokes_vel_eq} that the stochastic term is linked to 
the viscous term~\cite{LandauLifshtz1959} by the fluctuation-dissipation balance~\cite{FluctuationDissipation}.

\subsection{\label{subsec_spat_disc}Spatial Discretization}

\begin{figure}[t!]
\centering
\includegraphics[width=0.9\textwidth,clip]{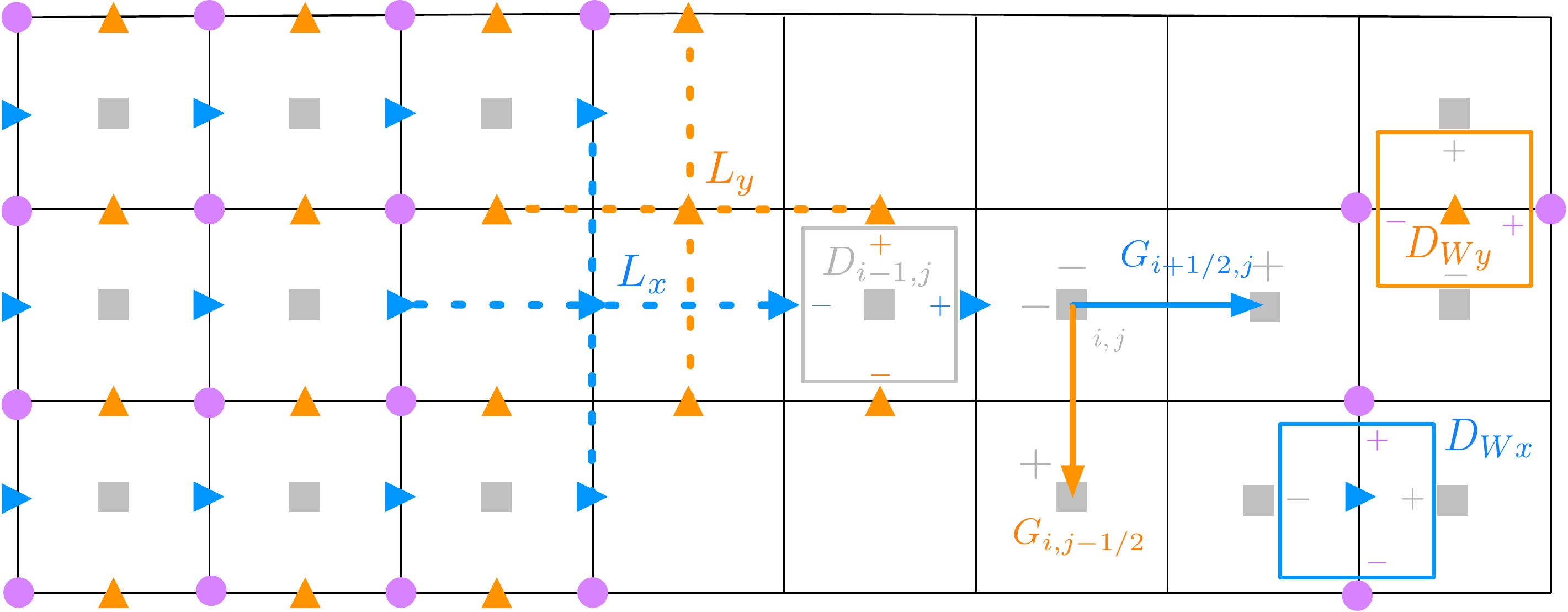}
\caption[]{\label{fig:MAC}
    Data layout for the two-dimensional periodic case.
    The velocity components ($U_x$ and $U_y$) are stored at the faces
    (\includegraphics[width=0.01\textwidth]{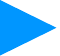} and 
    \includegraphics[width=0.01\textwidth]{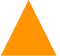} respectively).
    The pressure and Hodge variable components are stored at the cell centers 
    (\includegraphics[width=0.01\textwidth]{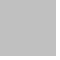}).
    The diagonal components of the stochastic stress ($W_{xx}$ and $W_{yy}$) are 
    stored at the cell centers, whereas the off-diagonal components 
    ($W_{xy}$ and $W_{yx}$) are stored at the nodes 
    (\includegraphics[width=0.01\textwidth]{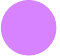}).
    All discrete differential operators are defined using the standard second-order
    centered finite differences.}
\end{figure}

To spatially discretize the stochastic incompressible Stokes equation 
\eqref{Stokes_vel_eq}, we employ the staggered-grid 
discretization on uniform mesh developed in~\cite{BalboaUsabiagaBellDelgadoBuscalioniDonvFaiGriffithPeskin2012}. 
While we consider the two-dimensional case in this paper for clarity,
the three-dimensional case is essentially the same.
We denote the spatially discretized equation as
\begin{equation}
\label{disc_vel_eq}
    \frac{dU}{dt} = {P}\left[ \nu L U 
    + \sqrt{\frac{2k_\mathrm{B}T\nu}{\rho\Delta V}} D_W W \right],
\end{equation}
where $U$ and $W$ are respectively the discretizations of $\mathbf{u}$ and 
the stochastic tensor field $\boldsymbol{\mathcal{W}}$;
${P}$ and $L$ are respectively the discrete projection and vector Laplacian operators; 
$D_W$ is the discrete divergence operator acting on $W$;
and $\Delta V=\Delta x \times\Delta y$ is the volume of a cell.
The data layout and discrete operators are illustrated in Figure~\ref{fig:MAC} and described next.
In section~\ref{subsubsec_vars_ops}, we first describe variables and operators,
assuming periodic boundary conditions.
In section~\ref{subsubsec_boundary_cond}, we explain modifications needed
to impose the no-slip boundary condition.

\subsubsection{\label{subsubsec_vars_ops}Variables and Operators}

According to the marker-and-cell layout~\cite{HarlowWelch1965}, 
the velocity components ($U_x$ and $U_y$) are stored at the faces, 
whereas the pressure and Hodge variable (denoted as $\Phi$) components are 
stored at the cell centers.
The discrete divergence operator $D$ and discrete gradient operator $G$ are defined
as follows. 
At a cell center $c$, the divergence of the velocity, $(DU)_c$, is constructed 
using the second-order centered finite difference.
At face $f$, the Hodge gradient $(G\Phi)_f$ is also computed 
using the second-order centered finite difference.
With these definitions, desired relations that hold in the continuous case are 
still valid.
First, the discrete gradient and divergence operators obey the duality relation 
$D^* = -G$.
From the definition of the discrete projection operator $P=I-G(DG)^{-1}D$, 
it can be easily seen that $P$ is indeed an orthogonal projection,
\ie $P^2=P$ and $P^*=P$.\footnote{The duality relation between the discrete gradient and divergence operators is essential for the projection property and ultimately for the stability of the method. The relation is easily satisfied here because the grid is uniform. On adaptive grids, it requires careful discretization (see \eg \cite{GUITTET2015,THEILLARD201991,THEILLARD2019108841}) or adequate function basis selection (see \eg \cite{PLUNKETT2014121}).}
The discrete Laplacian of the velocity at face $f$, denoted by $(LU)_f$, is 
computed using the standard second-order 5-point stencil.

The diagonal components of the stochastic stress tensor ($W_{xx}$ and $W_{yy}$) are 
stored at the cell centers, while the off-diagonal components ($W_{xy}$ and $W_{yx}$)
are stored at the nodes of the mesh (see Figure~\ref{fig:MAC}).
These stochastic terms are constructed as follows.
Since each component of the stochastic noise $\mathcal{W}_{ij}$ is a distribution, 
it cannot simply be evaluated at any given point and must be interpreted in the integral form.
That is, its discretization $\overline{W}_{ij}$ is constructed as 
the spatial average over the volume $V_{ij}$ of size $\Delta V$, 
centered where $\overline{W}_{ij}$ is stored:
\begin{equation}
  \overline{W}_{ij}(t) = \frac{1}{\Delta V} 
  \int_{V_{ij}} \mathcal{W}_{ij}(\mathbf{r}, t) \mathrm{d}\mathbf{r}.
\end{equation}
Hence, each component $\overline{W}_{ij}$ defined at each cell center or node is 
an independent Gaussian white noise process with variance $\Delta V^{-1}$.
To explicitly express the dependence of the magnitude of fluctuations on $\Delta V$,
we introduce normalized stochastic processes $W(t)=\overline{W}(t)/\sqrt{\Delta V}$.
The covariance of $W(t)$ is expressed as
\begin{equation}
\label{Cov_W_disc}
    \langle W(t) W(t') \rangle = C_W \delta(t-t').
\end{equation}
While $C_W$ is simply the identity matrix in the periodic boundary case, 
we will see that $C_W$ needs to be modified for the no-slip boundary condition.

One of the crucial issues for spatial discretization is that the discrete system 
should reproduce a correct equilibrium distribution that is expected from 
the continuous case.
Since the fluctuation-dissipation balance principle dictates the equilibrium 
in the continuous case, it is required that its discrete version should 
be satisfied when a spatial discretization is constructed~\cite{ATZBERGER20071255,PAZNER2019100068}.
In other words, the discrete fluctuation-dissipation balance dictates
the choice of $D_W$.
Since $D_W$ is a discretization of a divergence operator, it is natural to base 
its construct on $D$, keeping in mind that $D$ acting on $U$ is defined at the cell 
centers, while $D_W$ acting on $W$ is defined at the faces.
Using the second-order centered finite differences, 
the discrete stochastic divergence $D_W$ is constructed.
Then it can easily be seen that the following discrete fluctuation-dissipation balance
\begin{equation}
\label{FDB_disc}
    L = L^* = - D_W C_W D_W^*
\end{equation}
is satisfied.
The properties of the discretized operators $L$ and $P$ are important 
for computing the steady-state covariance in sections~\ref{sec:cov_and_struct_fac}
and \ref{sec:sscov}.

\subsubsection{\label{subsubsec_boundary_cond}Boundary Conditions}

\begin{figure}
\centering
\includegraphics[width=0.5\textwidth]{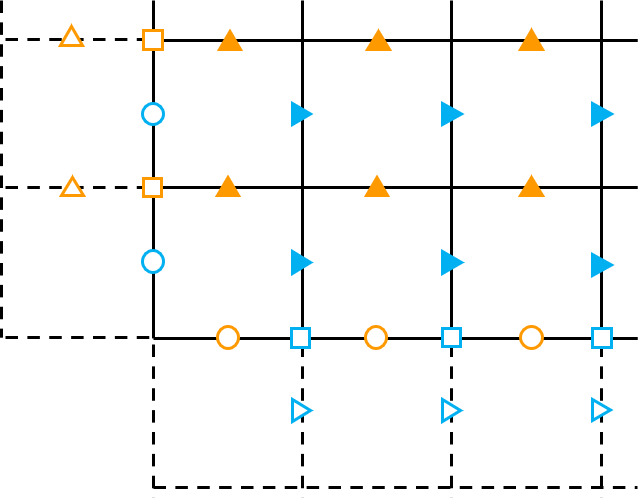}
\caption{\label{fig:BC}
    Data layout for the two-dimensional no-slip boundary case around the left-bottom corner.
    Velocity components contained in $U$ are denoted by filled triangles, 
    whereas empty triangles depict velocity components belonging to ghost cells.
    Normal velocity components at the boundaries, set to be zero, 
    are shown as empty circles.
    Empty squares denote tangential velocity components prescribed at the boundaries.}
\end{figure}

In the presence of non-periodic boundaries, the discrete operators defined above need to
be modified near the boundaries.
For the no-slip boundary condition, the velocity component normal to the boundary 
is zero at the boundary.
Hence, the velocity components at faces on the boundary are set to zero and not included as independent degrees of freedom (see Figure~\ref{fig:BC}).
Then, no values in cells outside the physical domain (\ie ghost cells) are needed to define $D$.
For the projection step, this zero normal velocity condition implies that
the homogeneous Neumann boundary condition should be chosen for the Hodge variable~\cite{ELiu1995}.
Hence, ghost cell values for the Hodge variable are set to be equal to the values 
in the neighboring interior cells, and the resulting $G$ operator satisfies 
the duality relation with
$D$~\cite{BalboaUsabiagaBellDelgadoBuscalioniDonvFaiGriffithPeskin2012}.
Therefore, $P^2=P$ and $P^*=P$ continue to hold for $P=I-G(DG)^{-1}D$.

To define the discrete vector Laplacian $L$, ghost cells are needed for
parallel velocity components that are half a grid spacing away
from the boundary (see Figure~\ref{fig:BC}).
When the tangential velocity is prescribed at the nodes on the boundary,
the corresponding ghost cell value is determined 
by the linear extrapolation~\cite{BalboaUsabiagaBellDelgadoBuscalioniDonvFaiGriffithPeskin2012}.
Then, $L$ can be constructed using the 5-point stencil, and $LU$ can be expressed as
\begin{equation}
\label{LL0B}
    L U = L_0 U + B \bar{U},
\end{equation}
where $\bar{U}$ denotes the prescribed tangential velocity.
Since we consider the homogeneous Dirichlet boundary condition (\ie $\bar{U}=0$),
$LU$ becomes $L_0 U$, and thus $B$ does not appear.
However, it is noted that the intermediate velocity appearing in a projection method
may not satisfy the prescribed boundary condition.
To develop and analyze our numerical schemes, we will use
this representation for nonzero tangential velocities.
We note that the size of $B$ depends on how $\bar{U}$ is represented.
For the data layout of $\bar{U}$, we use the same one as $U$, where each tangential
velocity prescribed on the boundaries (empty squares in Figure~\ref{fig:BC}) is stored at the location of the face (the closest filled triangle) half a grid spacing inward from the actual location of the prescribed velocity on the boundaries.
In this setting, the size of $B$ is identical to that of $L_0$.

Finally, since $L$ is modified, $C_W$ also needs to be modified to satisfy 
the discrete fluctuation-dissipation balance \eqref{FDB_disc}.
This can be achieved by setting the variance of stochastic fluxes $W_{ij}$
(affecting the tangential velocity components) at nodes on the boundary 
to twice that used for the interior
fluxes~\cite{BalboaUsabiagaBellDelgadoBuscalioniDonvFaiGriffithPeskin2012}.
In other words, $C_W$ is still a diagonal matrix with most diagonal elements being one but 
diagonal elements corresponding to those nodes on the boundary become two.

\subsection{\label{subsec_temp_disc}Temporal Discretization}

Following \cite{DelongGriffithVandenEijndenDonev2013}, we discretize \eqref{disc_vel_eq} in time as
\begin{equation}
  \label{CNupdate}
  U^{n+1} = P\left[U^n + \nu \Delta t L\left(\frac{U^{n} + U^{n+1}}{2}\right) 
  + \sqrt{\frac{2 k_\mathrm{B}T \nu \Delta t}{\rho\Delta V}} D_W W^{n}\right].
\end{equation}
Here, superscripts denote timesteps.
Since the covariance of $\int_{t^n}^{t^n+\Delta t} W(t)dt$
is proportional to $\Delta t$, 
the collection of the Gaussian white noise processes $W(t)$ has been replaced 
by $\sqrt{\Delta t} W^n$, where each component of $W^n$ defined 
at each spatial point at each time step is an independent standard normal random 
variate.
As explained in section~\ref{subsubsec_boundary_cond}, the variance of $W^n$
needs to be doubled on the no-slip boundary to satisfy the discrete
fluctuation-dissipation balance~\eqref{FDB_disc}.

This scheme is obtained from the Crank--Nicolson approximation with the assumption
$PU^n = U^n$.
Note that the resulting $U^{n+1}$ also stays in the projected space.
As explained in section~\ref{subsec_cov_CN}, this scheme does not introduce
any time discretization error in the equilibrium covariance
(or equivalently, the equilibrium structure factor)~\cite{DelongGriffithVandenEijndenDonev2013}.
In order to implement this scheme, however,
a linear solver~\cite{CaiNonakaBellGriffithDonev2014} for the following coupled 
system is required~\cite{DelongGriffithVandenEijndenDonev2013}:
\begin{subequations}
\label{saddle_point_lin_sys}
\begin{equation}
\label{saddle_point_lin_sys1}
    \left(I-\frac12\nu\Delta t L \right) U^{n+1} + \Delta t G\Pi^{n+\frac12}
    = \left(I+\frac12\nu\Delta t L \right) U^n 
    + \sqrt{\frac{2 k_\mathrm{B} T \nu\Delta t}{\rho\Delta V}} D_W W^n,
\end{equation}
\begin{equation}
    \label{saddle_point_lin_sys2}
    D U^{n+1} = 0.
\end{equation}
\end{subequations}
The main goal of this paper is to solve \eqref{CNupdate} not using
a linear solver for the saddle-point system~\eqref{saddle_point_lin_sys}
but using a projection method.


\section{\label{sec:cov_and_struct_fac}Steady-State Covariance and Static Structure Factor}

In this section, we introduce quantities that characterize the behavior of fluctuations
in the equilibrium Stokes system and corresponding discretized systems.
Since the mean velocity is zero at equilibrium, we focus on the second moment.
The steady-state covariance measures the covariance of velocities at two physical locations
for a system in equilibrium.\footnote{One can also consider time-correlation functions, \ie $\mathcal{C}(\tau)=\lim_{t\rightarrow\infty}\langle\mathbf{u}(t)\mathbf{u}^*(t+\tau)\rangle$, and their space-time spectra, so-called dynamic structure factor~\cite{DonevVandenEijndenGarciaBell2010}. The dynamic structure factor gives more detailed information on the time evolution of the system.
In the present paper, however, we focus the (semi-)analytic analysis of the static structure factor.}
Since the velocity fields are represented by $\mathbf{u}(t)$, $U(t)$, and $U^n$ 
for the continuum, spatially discretized, and fully discretized cases, respectively,
the corresponding steady-state covariances are defined as
\begin{subequations}
\begin{eqnarray}
    \mbox{Continuum} \quad &\mathcal{R} & = \lim_{t\rightarrow\infty} \langle \mathbf{u}(t)\mathbf{u}^*(t) \rangle,\\
    \mbox{Spatially discretized} \quad &R_{\Delta \mathbf{x}}& = \lim_{t\rightarrow\infty} \langle U(t) U^*(t) \rangle,\\
    \mbox{Fully discretized} \quad &R_{\Delta \mathbf{x},\Delta t}& = \lim_{n\rightarrow\infty} \langle U^n (U^n)^* \rangle.
\end{eqnarray}
\end{subequations}
For each case, the static structure factor is defined as the Fourier transform
of the steady-state covariance.
We note that, for the no-slip boundary case, the velocity field is mirrored and 
the resulting field in the extended domain is used (see Figure~\ref{figappendix})
since the velocity field in the original domain is not periodic.

These quantities can be used to investigate
the accuracy of spatial discretization and numerical schemes.
In section~\ref{sec:sscov}, we analytically investigate our projection-method-based schemes by computing the steady-state covariance.
In section~\ref{sec_num_val}, we numerically investigate
those schemes by computing the static structure factor.
In this section, we derive the main analytic results for the continuum and spatially discretized cases
(\ie $\mathcal{R}$ and $R_{\Delta \mathbf{x}}$)
as well as the Crank--Nicolson scheme \eqref{CNupdate} 
(\ie $R_{\Delta \mathbf{x},\Delta t}^\mathrm{CN}$).
While those results are known~\cite{BalboaUsabiagaBellDelgadoBuscalioniDonvFaiGriffithPeskin2012}, 
we present them along with derivations, as both aspects are essential 
for the analysis of our projection-method-based schemes.
In \ref{appendix_proj_sdes}, we develop a systematic procedure to construct and solve
linear systems from which the steady-state covariance can be uniquely determined.

\subsection{Continuum and Spatially Discretized Cases}

Using the result of \ref{appendix_proj_sdes}, the steady-state covariance
$\mathcal{R}$ 
of the continuum equation~\eqref{Stokes_vel_eq} is given as the unique solution of
\begin{subequations}
\label{cond_cov_continuum}
\begin{equation}
\nu\mathcal{P}\mathcal{L}\mathcal{R}
+\nu\mathcal{R}\mathcal{L}\mathcal{P}
= -\frac{2k_\mathrm{B}T\nu}{\rho}\mathcal{P}\mathcal{D}\mathcal{D}^*\mathcal{P},
\end{equation}
\begin{equation}
\mathcal{P}\mathcal{R} = \mathcal{R}\mathcal{P} = \mathcal{R}.
\end{equation}
\end{subequations}
The physical intuition that the equilibrium covariance must be proportional to
the projection operator, suggests that
\begin{equation}
\label{R_conti}
\mathcal{R} = \frac{k_\mathrm{B}T}{\rho}\mathcal{P}.
\end{equation}
Using the properties of $\mathcal{P}$ and $\mathcal{L}$ mentioned
in section~\ref{subsec_conti_eqn} (\ie $\mathcal{P}^*=\mathcal{P}$,
$\mathcal{P}^2=\mathcal{P}$, $\mathcal{L}^*=\mathcal{L}$,
$\mathcal{L}=-\mathcal{D}\mathcal{D}^*$), it can be easily shown that
\eqref{R_conti} indeed satisfies \eqref{cond_cov_continuum}.

The static structure factor is obtained from the Fourier transform of
the steady-state covariance 
(with normalization~\cite{DonevVandenEijndenGarciaBell2010} with respect to the volume of the system $V$):
\begin{equation}
  \label{S_conti}
  \mathcal{S} = V \langle \widehat{\mathbf{u}}(t)\widehat{\mathbf{u}}^*(t) \rangle
  = \frac{k_\mathrm{B}T}{\rho}\widehat{\mathcal{P}},
\end{equation}
where a hat denotes the Fourier transform.
For the periodic boundary case, 
the structure factor has a compact form
\begin{equation}
  \label{S_pbc}
  \mathcal{S}(\mathbf{k}) = \frac{k_\mathrm{B}T}{\rho}
  \left[I - \frac{\mathbf{k}\mathbf{k}^*}{\mathbf{k}\cdot\mathbf{k}}\right].
\end{equation}
It is easy to see from \eqref{S_pbc} that all divergence-free modes have the same spectral power 
at equilibrium in the periodic boundary case.
In fact, we note that \eqref{S_conti} implies the same conclusion
for a general case where aforementioned properties of $\mathcal{P}$ and
$\mathcal{L}$ hold~\cite{BalboaUsabiagaBellDelgadoBuscalioniDonvFaiGriffithPeskin2012}.

For the spatially discretized equation~\eqref{disc_vel_eq}, a similar argument
can be made since the discrete operators $P$ and $L$ are constructed 
so that they preserve all aforementioned properties of $\mathcal{P}$ and $\mathcal{L}$.
Because in this case the fluctuation-dissipation balance is given in the form of \eqref{FDB_disc}, 
the linear system that uniquely determines $R_{\Delta \mathbf{x}}$ is
\begin{subequations}
\begin{equation}
    \nu P L R_{\Delta \mathbf{x}} + \nu R_{\Delta \mathbf{x}} L P 
    = -\frac{2k_\mathrm{B}T\nu}{\rho \Delta V} P D_W C_W D_W^* P
\end{equation}
\begin{equation}
    P R_{\Delta \mathbf{x}} = R_{\Delta \mathbf{x}} P = R_{\Delta \mathbf{x}}
\end{equation}
\end{subequations}
and, similarly to the continuum steady-state expression \eqref{R_conti}, the discrete steady-state covariance is given as
\begin{equation}
\label{R_disc}
  R_{\Delta \mathbf{x}} = \frac{k_\mathrm{B}T}{\rho\Delta V} P.
\end{equation}

\subsection{\label{subsec_cov_CN}Crank--Nicolson Scheme}

Introducing $A_\pm = I \pm \frac12\nu\Delta t PL$,
the Crank--Nicolson scheme \eqref{CNupdate} can be written as
\begin{equation}
\label{CNupdate2}
  U^{n+1} = A^{-1}_- A_+ U^n 
  + \sqrt{\frac{2k_\mathrm{B}T\nu\Delta t}{\rho\Delta V}} A^{-1}_- P D_W W^n.
\end{equation}
Hence, using the result of \ref{appendix_proj_sdes}, the steady-state covariance
$R_{\Delta \mathbf{x},\Delta t}^\mathrm{CN} = \langle U^n (U^n)^* \rangle$ is
given as the unique solution of
\begin{subequations}
\label{lin_sys_RCN}
\begin{equation}
  (A^{-1}_- A_+) R_{\Delta \mathbf{x},\Delta t}^\mathrm{CN}
  \left( A^{-1}_- A_+ \right)^*
  - R_{\Delta \mathbf{x},\Delta t}^\mathrm{CN}
  = -\frac{2k_\mathrm{B}T\nu\Delta t}{\rho\Delta V} 
  (A_-^{-1}P D_W) C_W \left(A_-^{-1}P D_W\right)^*,
\end{equation}
\begin{equation}
  P R_{\Delta \mathbf{x},\Delta t}^\mathrm{CN}
  = R_{\Delta \mathbf{x},\Delta t}^\mathrm{CN} P
  = R_{\Delta \mathbf{x},\Delta t}^\mathrm{CN}.
\end{equation}
\end{subequations}
By observing that
\begin{equation}
  A_+ P A_+^* - A_- P A_-^* = 2\nu\Delta t PLP,
\end{equation}
and using the discrete fluctuation-dissipation balance~\eqref{FDB_disc},
one can show that
\begin{equation}
  \label{R_CN}
  R_{\Delta \mathbf{x},\Delta t}^\mathrm{CN}
  = \frac{k_\mathrm{B}T}{\rho\Delta V} P
\end{equation}
satisfies \eqref{lin_sys_RCN}.
The results \eqref{R_disc} and \eqref{R_CN} show that the Crank--Nicolson scheme does not 
introduce any temporal integration errors to the steady-state covariance.


\section{\label{sec_construct_methods}Construction of Projection Methods}

We present here how our projection methods are constructed to compute the numerical solution
of the (spatially discretized) stochastic incompressible Stokes equation~\eqref{disc_vel_eq}.
The resulting schemes for the periodic boundary and no-slip boundary cases are given in Schemes~\ref{scheme_pbc} and \ref{scheme_K}, respectively.
Our projection-method-based schemes solve the Crank--Nicolson update~\eqref{CNupdate}
using the operator splitting approach.
In section~\ref{subsec_pbc_scheme}, we consider the periodic boundary case, 
where the simple splitting exactly solves \eqref{CNupdate}.
In section~\ref{subsec_hypo_scheme}, we first discuss issues that arise 
when the simple splitting is applied to the no-slip case and 
introduce the idea of iterative boundary corrections.
In section~\ref{subsec_Kiter_scheme}, we present our iterative scheme for the no-slip case.

\begin{algorithm}
\caption{\label{scheme_pbc}Projection method for the periodic boundary case}
    Given velocity $U^n$ at timestep $t^n$, the velocity $U^{n+1}$ at the next timestep 
    is updated via $\tilde{U}$ as follows.
    \begin{equation}
        \tilde{U} = U^n + \nu \Delta t L\left(\frac{U^{n} + \tilde{U}}{2}\right) 
        + \sqrt{\frac{2 k_\mathrm{B}T \nu \Delta t}{\rho\Delta V}} D_W W^{n},
    \end{equation}
    \begin{equation}
        DG\Phi = D\tilde{U},
    \end{equation}
    \begin{equation}
        U^{n+1} = \tilde{U}-G\Phi.
    \end{equation}
\end{algorithm}

\begin{algorithm}
\caption{\label{scheme_K} $K$-Iteration scheme for the no-slip boundary case $(1\le K<\infty)$}
    Given velocity $U^n$ at timestep $t^n$, the velocity $U^{n+1}$ at the next timestep is
    updated as follows.
    \begin{enumerate}
    \item For $k=1,2,\cdots,K$:
        \begin{itemize}
        \item Compute $\tilde{U}_k$ using $\Phi_{k-1}$. For $k=1$, use $\Phi_0 = 0$.
            \begin{equation}
            \tilde{U}_k = U^n + \nu\Delta t L_0\left(\frac{U^n+\tilde{U}_k}{2}\right)
            +\sqrt{\frac{2k_\mathrm{B}T\nu\Delta t}{\rho\Delta V}}D_W W^n
            +\frac12\nu\Delta t BG\Phi_{k-1}.
        \end{equation}
        \item Compute $\Phi_k$.
            \begin{equation}
            DG\Phi_k=D\tilde{U}_k.
            \end{equation}
        \end{itemize}
    \item Compute $U^{n+1}$ from the projection of $\tilde{U}_K$:
        \begin{equation}
        U^{n+1} = \tilde{U}_K - G\Phi_K.
        \end{equation}
\end{enumerate}
\end{algorithm}

\subsection{\label{subsec_pbc_scheme}Periodic Boundary Case}

For periodic boundary conditions, the Crank--Nicolson update \eqref{CNupdate} can be
exactly implemented by the simple projection-based time-splitting 
given in Scheme~\ref{scheme_pbc}.
Since both $\tilde{U}$ and $\Phi$ obey periodic boundary conditions,
$U^{n+1}$ also satisfies periodic boundary conditions.

Using the fact that $L$ and $P$ commute in the periodic boundary case,
we can show that Scheme~\ref{scheme_pbc} does not introduce any splitting error
to solve \eqref{CNupdate}.
We first observe that Scheme~\ref{scheme_pbc} can be written as
\begin{equation}
\label{projpbc2}
  U^{n+1} = P\tilde{U} 
  = P\left[U^n + \nu \Delta t L\left(\frac{U^{n} + \tilde{U}}{2}\right) 
  + \sqrt{\frac{2 k_\mathrm{B}T \nu \Delta t}{\rho\Delta V}} D_W W^{n}\right].
\end{equation}
Using the commutativity of $L$ and $P$, we have
$PL\tilde{U} = PPL\tilde{U} = PLP\tilde{U} = PLU^{n+1}$ 
and thus see that \eqref{projpbc2} is identical to \eqref{CNupdate}.

\subsection{\label{subsec_hypo_scheme}No-Slip Boundary Case: Hypothetical Scheme}

When Scheme~\ref{scheme_pbc} is applied to the no-slip boundary case 
with $L$ replaced by $L_0$ (see \eqref{LL0B}), it does not exactly solve \eqref{CNupdate}.
Since $\tilde{U}$ is solved with the homogeneous Dirichlet boundary condition and 
$\Phi$ is solved with the homogeneous Neumann boundary condition, 
the resulting $U^{n+1}=\tilde{U}-G\Phi$ does not necessarily satisfy 
the homogeneous Dirichlet boundary condition.
While its normal component is zero, its parallel component is not guaranteed to be zero.

If we \textit{could} impose the Dirichlet boundary condition $\tilde{U}=G\Phi$ 
to $\tilde{U}$, the resulting $U^{n+1}=\tilde{U}-G\Phi$ \textit{would} satisfy
the homogeneous Dirichlet boundary condition.
By expressing the vector Laplacian operator $L$ with a specified Dirichlet boundary
condition in terms of $L_0$ and $B$ (see \eqref{LL0B}), 
this procedure can be written as
\begin{subequations}
\label{hypo_scheme}
    \begin{equation}
    \label{hypo_scheme_step1}
        \tilde{U} = U^n + \nu\Delta t L_0\left(\frac{U^n+\tilde{U}}{2}\right)
        +\sqrt{\frac{2k_\mathrm{B}T\nu\Delta t}{\rho\Delta V}}D_W W^n
        +\frac12\nu\Delta t BG\Phi,
    \end{equation}
    \begin{equation}
    \label{hypo_scheme_step2}
        DG\Phi=D\tilde{U},    
    \end{equation}
    \begin{equation}
        \label{hypo_scheme_step3}
        U^{n+1}=\tilde{U}-G\Phi.
    \end{equation}
\end{subequations}
However, the first two steps in this scheme cannot be sequentially implemented 
because $G\Phi$ is not available when $\tilde{U}$ is computed in \eqref{hypo_scheme_step1}
and only available after \eqref{hypo_scheme_step2}.
Nonetheless, this \textit{hypothetical} scheme is worth investigating
because it exactly solves the Crank--Nicolson update~\eqref{CNupdate}.

We show that \eqref{hypo_scheme} is equivalent to \eqref{CNupdate}
by using the commutativity of $L_0+B$ and $P$, \ie
\begin{equation}
\label{commutLP}
    P(L_0+B) = (L_0+B)P.
\end{equation}
For the proof of \eqref{commutLP}, see \ref{appendix_proof_comm}.
By applying $P$ to \eqref{hypo_scheme_step1} and using $U^{n+1}=P\tilde{U}$ and
$G\Phi = Q\tilde{U}$ where $Q = I - P$, we obtain
\begin{equation}
\label{hypo_scheme2}
    U^{n+1} -\frac12\nu\Delta t\left[ P L_0 \tilde{U} + PBQ\tilde{U}\right]
    = P\left[ U^n + \frac12\nu\Delta t L_0 U^n
    +\sqrt{\frac{2k_\mathrm{B}T\nu\Delta t}{\rho\Delta V}}D_W W^n \right].
\end{equation}
Since it can be shown using \eqref{commutLP} that
\begin{equation}
  PL_0\tilde{U} = PPL_0\tilde{U}
  = P\left(L_0P\tilde{U} + BP\tilde{U} - PB\tilde{U}\right)
  = PL_0 U^{n+1} - PBQ\tilde{U},
\end{equation}
\eqref{hypo_scheme2} becomes identical to \eqref{CNupdate} and 
therefore the hypothetical scheme~\eqref{hypo_scheme} would not introduce any splitting error.

\subsection{\label{subsec_Kiter_scheme}
No-Slip Boundary Case: $K$-Iteration Scheme $(1\le K < \infty)$}

To construct implementable schemes based on the hypothetical scheme~\eqref{hypo_scheme},
we first observe that $\tilde{U}$ satisfies the following fixed-point problem:
\begin{equation}
  \label{fixedprob}
  \tilde{U} = \left(I-\frac12\nu\Delta t L_0\right)^{-1}
  \left[\left(I+\frac12\nu\Delta t L_0\right)U^n+\frac12\nu\Delta t BQ\tilde{U}
  + \sqrt{\frac{2k_\mathrm{B}T\nu\Delta t}{\rho\Delta V}}D_W W^n
  \right]
  \equiv\Psi(\tilde{U}).
\end{equation}
We then construct the following iteration procedure 
of computing $\tilde{U}_1$, $\Phi_1$, $\tilde{U}_2$, $\Phi_2$, $\cdots$
to obtain the convergent solution $\tilde{U}_\infty = \lim_{k\rightarrow\infty} \tilde{U}_k$ 
and $\Phi_\infty = \lim_{k\rightarrow\infty} \Phi_k$:
\begin{subequations}
\label{iterationsonly}
\begin{equation}
\label{iter1}
  \tilde{U}_k = U^n + \nu\Delta t L_0\left(\frac{U^n+\tilde{U}_k}{2}\right)
  +\sqrt{\frac{2k_\mathrm{B}T\nu\Delta t}{\rho\Delta V}}D_W W^n
  +\frac12\nu\Delta t BG\Phi_{k-1},
\end{equation}
\begin{equation}
\label{iter2}
  DG\Phi_k=D\tilde{U}_k,
\end{equation}
\end{subequations}
where we assume $\Phi_0 = 0$.
Since $G\Phi_{k-1}=Q\tilde{U}_{k-1}$, it is easy to see that, if \eqref{iterationsonly} converges, 
the limit $\tilde{U}_\infty$ is the solution of the fixed-point problem \eqref{fixedprob} 
and, equivalently, the solution $\tilde{U}$ of \eqref{hypo_scheme_step1} and \eqref{hypo_scheme_step2} 
in the hypothetical scheme.
Therefore, the no-slip boundary solution for the next timestep is obtained as 
$U^{n+1}=\tilde{U}_\infty - G\Phi_\infty$. 
We also note that the last term in \eqref{iter1} is the boundary condition correction 
using $\tilde{U}_{k-1}$. 

Since the iteration \eqref{iterationsonly} can be written as $\tilde{U}_k = M\tilde{U}_{k-1} + c$,
where
\begin{equation}
    M = \frac12\nu\Delta t\left(I-\frac12\nu\Delta t L_0\right)^{-1} B Q,
\end{equation}
the convergence criterion is that the spectral radius $\rho(M)$ of $M$ is smaller than 1.
The rate of convergence $r$ (\ie $\lVert\tilde{U}_k-\tilde{U}\rVert \sim r^k$) is given as
\begin{equation}
\label{spectrad}
    r = \rho(M).
\end{equation}
Finally, by specifying a finite number $K$ of iterations ($K=1,2,\cdots)$, 
we obtain Scheme~\ref{scheme_K}, which we call the $K$-iteration scheme.
Note that the 1-iteration scheme corresponds to the projection method
where no boundary correction is considered.
Alternatively, one can impose a convergence criterion such as
\begin{equation}
\label{eq:term_crit}
  \frac{\lVert\Phi_{k-1}-\Phi_k\rVert}{\lVert\Phi_k\rVert} \le \epsilon.
\end{equation}


\section{\label{sec:sscov}Steady-State Covariance Error Analysis}

In this section, we analyze the accuracy of the $K$-iteration scheme
(see Scheme~\ref{scheme_K}) 
by investigating the resulting steady-state covariance matrix 
$R^{(K)}_{\Delta \mathbf{x},\Delta t} = \lim_{n\rightarrow\infty}\langle U^n (U^n)^*\rangle$.
While the same results can be obtained using the structure factor 
(note that the structure factor is basically the Fourier transform of 
the steady-state covariance), we use the steady-state covariance in this section; 
we analyze the structure factor in section~\ref{sec_num_val}, 
where numerical validation results are presented.

As shown in section~\ref{subsec_hypo_scheme}, if infinite iterations were performed 
each timestep, the resulting $\infty$-iteration projection method would give
the identical temporal update as the Crank--Nicolson scheme and thus 
its steady-state covariance would not have any temporal integration errors
(\ie $R^{(\infty)}_{\Delta \mathbf{x},\Delta t} = R_{\Delta \mathbf{x}}$).
When a finite number $K$ of iterations are used, 
the new state $U^{n+1}=P\tilde{U}_K$ is computed from inexact $\tilde{U}_K$,
causing temporal integration errors in $R^{(K)}_{\Delta \mathbf{x},\Delta t}$.
The main theoretical result of this section is that the temporal integration error of 
the $K$-iteration scheme in the steady-state covariance is of the order of $\Delta t^K$:
\begin{equation}
\label{RKRx}
    R^{(K)}_{\Delta \mathbf{x},\Delta t} - R_{\Delta \mathbf{x}} = O(\Delta t^K).
\end{equation}
In section~\ref{subsec_anal_res}, we show that temporal integration errors committed 
at each timestep due to a finite number of boundary corrections are $O(\Delta t ^K)$:
\begin{equation}
\label{UKUinfbrief}
    \tilde{U}_K - \tilde{U}_\infty = O(\Delta t^K).
\end{equation}
We note that this result strongly supports \eqref{RKRx}.
In section~\ref{sec:semianalytic}, we confirm \eqref{RKRx} using a semi-analytic approach.
That is, by noting that $R^{(K)}_{\Delta \mathbf{x},\Delta t}$ can be determined as the unique solution of a linear system described in \ref{appendix_proj_sdes},
we directly compute it for specific values of $K$ and $\Delta t$ for some small-sized systems
by solving the linear system.

\subsection{\label{subsec_anal_res}Analytic Results}

To show \eqref{UKUinfbrief}, we first derive expressions of $A_K$ and $B_K$ so that the temporal update of the $K$-iteration scheme can be expressed in the compact form
\begin{equation}
\label{proj_time_update}
    U^{n+1}=P\tilde{U}_K = P\left[A_K U^n + B_K W^n\right].
\end{equation}
By introducing $A_\pm = I \pm \frac12\nu\Delta t L_0$, we express $\tilde{U}_1$ as
\begin{equation}
    \tilde{U}_1 = A_-^{-1} \left[A_+ U^n + \sqrtfacdVdt D_W W^n\right].
\end{equation}
Since $G\Phi_k = Q\tilde{U}_k$, where $Q=I-P$, we recursively express $\tilde{U}_k$ 
for $k=2,3,\dots$,
and obtain the following general expression:
\begin{equation}
    \tilde{U}_K = A_-^{-1} \left[\sum_{k=0}^{K-1}\left(\frac12\nu\Delta t BQ
    A_-^{-1}\right)^k\right]
    \left[A_+ U^n + \sqrtfacdVdt D_W W^n\right].
\end{equation}
Using the identity 
$A_-^{-1} \left[\sum_{k=0}^{K-1}\left(\frac12\nu\Delta t BQ A_-^{-1}\right)^k\right]
= \left[\sum_{k=0}^{K-1}\left(\frac12\nu\Delta t A_-^{-1} BQ \right)^k\right] A_-^{-1}$,
we have an alternative expression for $\tilde{U}_K$, which gives the following expressions 
for $A_K$ and $B_K$ in the temporal update \eqref{proj_time_update}: 
\begin{equation}
\label{AKBK}
    A_K = \left[\sum_{k=0}^{K-1}\left(\frac12\nu\Delta t A_-^{-1} BQ \right)^k\right] 
    A_-^{-1} A_+,\quad
    B_K = \sqrtfacdVdt
    \left[\sum_{k=0}^{K-1}\left(\frac12\nu\Delta t A_-^{-1} BQ \right)^k\right] A_-^{-1} D_W.
\end{equation}
Using the well-known result for the geometric series of a matrix, we finally obtain
\begin{equation}
\label{UKUinf}
    \tilde{U}_K = \left[I-\left(\frac12\nu\Delta t A_-^{-1}BQ\right)^K\right]\tilde{U}_\infty
\end{equation}
and thus \eqref{UKUinfbrief}.

We note that no notion of stochastic accuracy (\eg weak vs.\ strong) is required to interpret
\eqref{UKUinfbrief} or \eqref{UKUinf} since $U^n$ and $W^n$ are fixed 
during the boundary correction iterations.
However, to define the order of temporal integration errors committed at each timestep,
a notion of stochastic order of convergence is needed.
In this paper, instead of analyzing the weak or strong orders of accuracy of our schemes, we focus on the convergence of the resulting steady-state covariance 
(and equivalently, the structure factor).
For discussion of stochastic accuracy of FHD schemes, we refer the reader to~\cite{DelongGriffithVandenEijndenDonev2013}.

\subsection{\label{sec:semianalytic}Semi-Analytic Results}

As described in \ref{appendix_proj_sdes}, 
one can derive a linear system (\eqref{Cov_rel_proj2} and \eqref{Cov_rel_proj_disc}, 
or equivalently, \eqref{biglinsysdisc})
that uniquely determines the covariance matrix,
using the definition of discretized operators and the expressions of $A_K$ and $B_K$ 
(see \eqref{proj_time_update} and \eqref{AKBK}).
However, the solution cannot be given explicitly.
Hence, we construct the linear system for a spatially discretized system of a specific size and
compute its solution numerically.
Although the solution has numerical errors due to floating-point arithmetic, 
these errors can be controllable and kept small comparable to machine precision.
In this sense, this approach gives semi-analytic results, which should not be confused with
stochastic simulation results (given in section~\ref{sec_num_val}) containing sampling errors.

We present results obtained from a two-dimensional system with $10\times10$ cells, where
no-slip boundary conditions are imposed on all boundaries.
We note that the choice of the number of cells is arbitrary, and the conclusion 
should not change for moderate to large system sizes (roughly speaking, $N\gtrsim 8$) 
as we justify below.
However, we point out that semi-analytic results for a larger system quickly 
become computationally infeasible.
This is because the size of the extended linear system \eqref{biglinsysdisc}
has a matrix with $(N-1)^4\times(N-1)^4$ components for a system domain 
with $N\times N$ cells.
We assume $\Delta x =\Delta y$ and define the dimensionless number
\begin{equation}
\label{beta}
    \beta = \frac{\nu\Delta t}{\Delta x^2}
\end{equation}
and investigate how the errors change as the value of $\beta$ varies. 
For faces $f$ and $f'$, we define the error at the $(f,f')$-component as 
\begin{equation}
\label{errEff}
    E_{f,f'} = \Delta V \left(R^{(K)}_{\Delta\mathbf{x},\Delta t}\right)_{f,f'}
    -\frac{k_\mathrm{B}T}{\rho}P_{f,f'},
\end{equation}
and define the maximum error as
\begin{equation}
    \varepsilon_\mathrm{max} = \max_{(f,f')}\lvert E_{f,f'}\rvert.
\end{equation}
Since $R^{(K)}_{\Delta\mathbf{x},\Delta t}$ is linearly proportional to $k_\mathrm{B}T/\rho$, 
we simply set $k_\mathrm{B}T/\rho = 1$.

\begin{figure}
    \centering
    \includegraphics[width=0.95\linewidth]{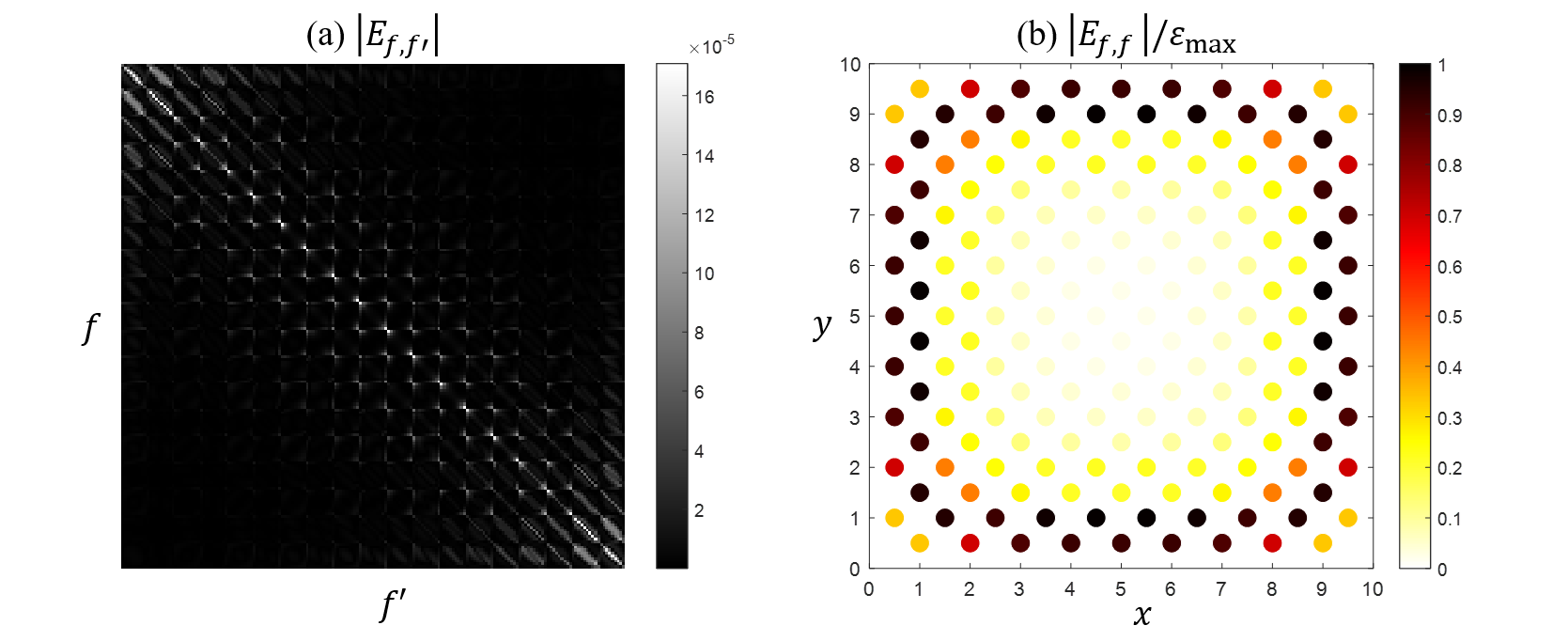}
    \caption{\label{fig3}
    Errors in $R^{(K)}_{\Delta\mathbf{x},\Delta t}$ for $K=2$ and $\beta=0.1$.
    In panel (a), each pixel represents a component of the error matrix $E$ and 
    its shade represents the value of $\lvert E_{f,f'}\rvert$.
    In panel (b), the physical domain of the system is shown and at each face $f$ the magnitude of the diagonal component $E_{f,f}$ (normalized by the maximum error $\varepsilon_\mathrm{max}$) is shown by the color map.}
\end{figure}

Figure~\ref{fig3} shows how the errors in the steady-state covariance matrix
$R^{(K)}_{\Delta\mathbf{x},\Delta t}$ are distributed for the 2-iteration scheme with $\beta=0.1$.
In panel~(a), the error matrix $E$ is displayed as a grayscale image where the brightness of a pixel represents the magnitude of each component $\lvert E_{f,f'}\rvert$.
The image shows that errors are dominant along the diagonal (\ie for $f=f'$).
This is because diagonal components of the steady-state covariance matrix tend to be larger than
off-diagonal components.
In panel~(b), the magnitude of each diagonal component $\lvert E_{f,f}\rvert$ is shown
at the corresponding face $f$ in the physical domain of the system.
The image indicates that errors are dominant near the boundaries.
This observation is consistent with the fact that the inexact boundary correction causes temporal integration errors of the $K$-iteration scheme
due to a finite number of iterations.
We note that the essentially same error pattern is observed for different system sizes.
For $N\times N$ cells with $8\le N\le 13$, the error distributions at the corners of the domain
and at the sides of the domain remain the same.
Moreover, compared with the maximum error $\varepsilon_\mathrm{max}$ for $N=10$, 
the corresponding values for $N=11$, 12, 13 have negligible deviations ($\lesssim$ 0.2\%).
Hence, we expect our observations for the $10\times 10$ system to remain valid for larger systems.

\begin{figure}
\centering
\includegraphics[width=0.33\textwidth,angle=270]{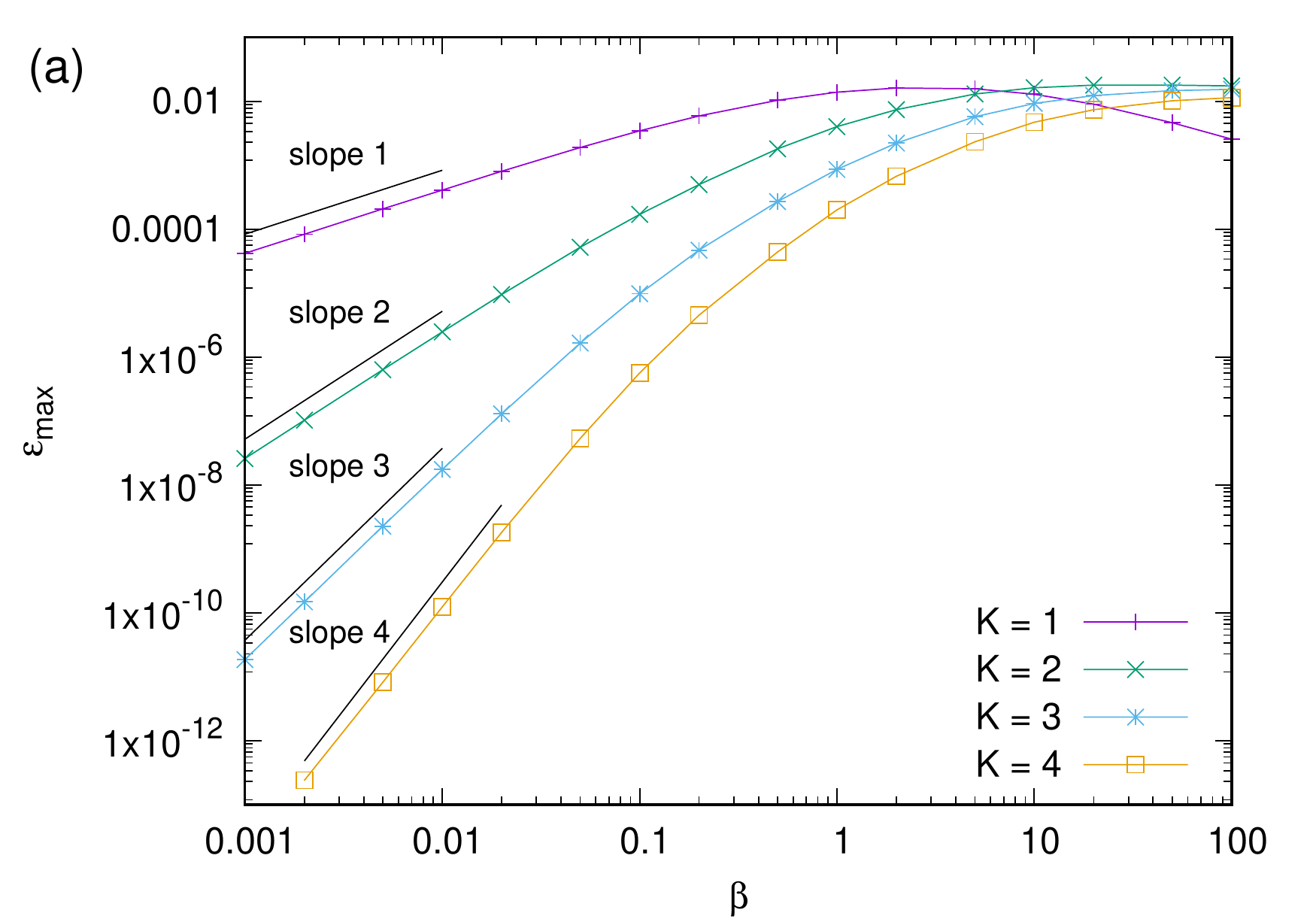}
\includegraphics[width=0.33\textwidth,angle=270]{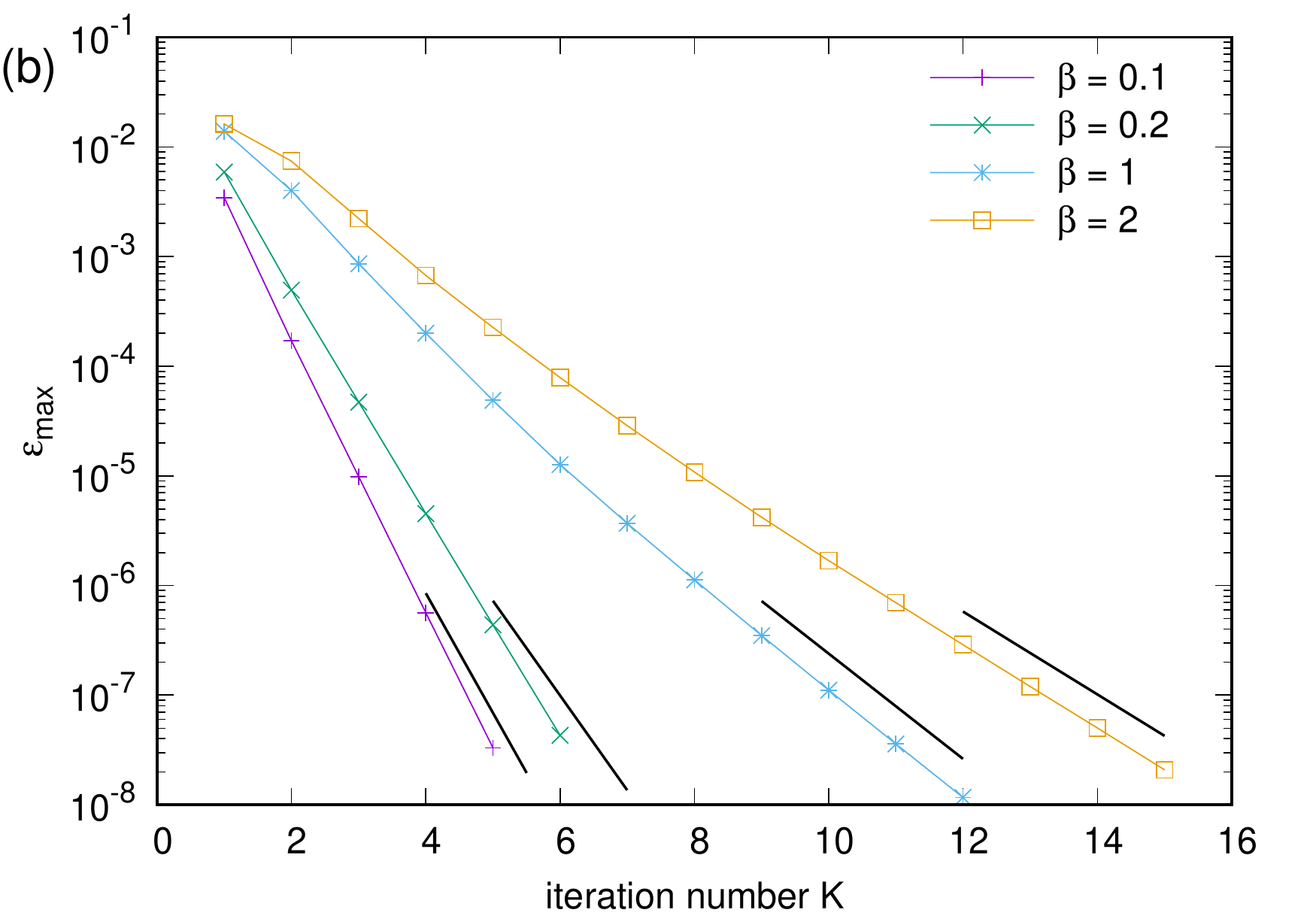}
\caption{\label{fig4}
    Dependence of the maximum error $\varepsilon_\mathrm{max}$ of
    $R^{(K)}_{\Delta\mathbf{x},\Delta t}$ on the number of iterations $K$ and 
    the stability condition number $\beta = \nu\Delta t/\Delta x^2$.
    In panel (a), for $K=1, 2, 3, 4,$ $\varepsilon_\mathrm{max}$ is plotted versus $\beta$ 
    in the log-log scales.
    Guide lines for integer slopes are shown for small $\beta$ values.
    In panel (b), $\varepsilon_\mathrm{max}$ is plotted versus $K$ 
    for various values of $\beta$ in the semi-log scales.
    The solid lines indicate the expected asymptotic decays given as $\mathrm{(const)}r^K$, 
    where $r$ is estimated from the spectral radius of the iteration matrix, 
    see \eqref{spectrad}.}
\end{figure}

In Figure~\ref{fig4}, we show the dependence of the maximum error $\varepsilon_\mathrm{max}$
on the number of iterations $K$ and the stability condition number $\beta$.
Figure~\ref{fig4}a demonstrates that the steady-state covariance matrix obtained 
by the $K$-iteration scheme has the $O(\Delta t^K)$ accuracy (see \eqref{RKRx}).
The semi-log plot of $\varepsilon_\mathrm{max}$ versus $K$ in Figure~\ref{fig4}b
indicates that for a given value of $\Delta t$ the error decreases exponentially 
with respect to $K$.
Moreover, we confirm that the errors asymptotically decay with the rate of convergence $r$ 
given in \eqref{spectrad} for large values of~$K$.


\section{\label{sec_num_val}Numerical Validations}

In this section, we present numerical validation results of our $K$-iteration scheme.
We solve the two-dimensional stochastic Stokes equation \eqref{Stokes_vel_eq} 
with no-slip boundary conditions using the $K$-iteration scheme
and calculate the equilibrium structure factors 
$S^{U_x}_{\Delta \mathbf{x},\Delta t}$ and $S^{U_y}_{\Delta \mathbf{x},\Delta t}$
using the time trajectories of $U^n = (U_x(n\Delta t),U_y(n\Delta t))$.
We compare these numerical results with the exact results for the discretized system
as well as the semi-analytical results for $\Delta t>0$ (available only for small systems).

\begin{figure}[t!]
\centering
\includegraphics[width=\textwidth]{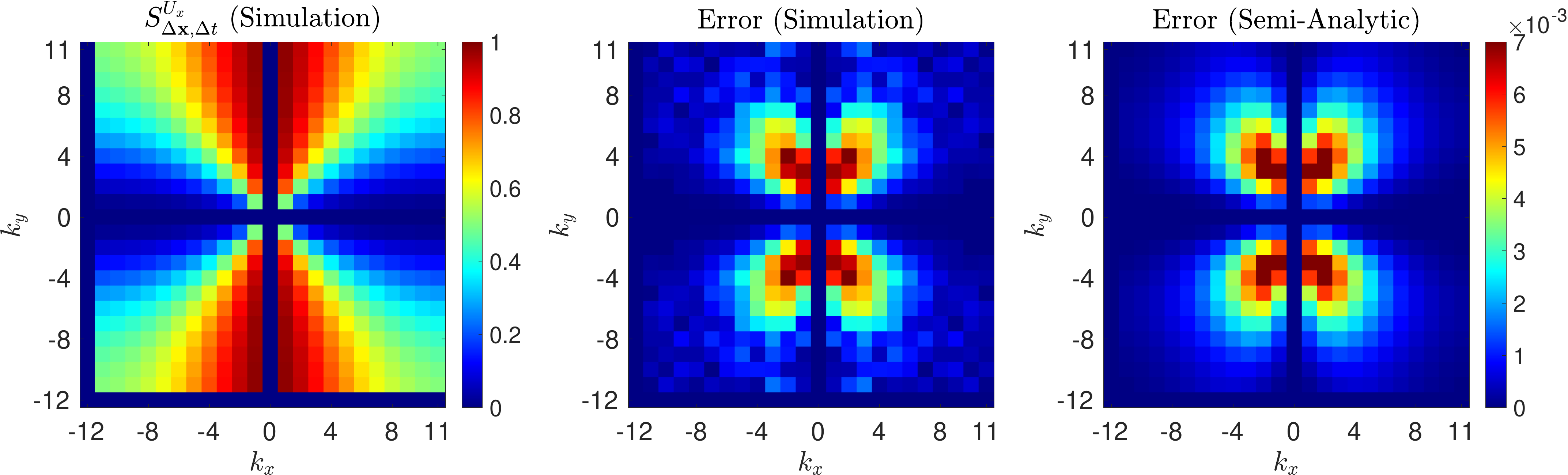}
\caption{\label{fig:structfact12x12}
    The equilibrium structure factor $S^{U_x}_{\Delta \mathbf{x},\Delta t}$ computed
    using the 2-iteration scheme (\ie $K=2$).
    The system has $12\times12$ cells and the stability condition number is 
    $\beta = \nu\Delta t/\Delta x^2 = 1$.
    The left panel displays $S^{U_x}_{\Delta \mathbf{x},\Delta t}$ obtained 
    from trajectory calculation up to $N_2=10^7$ timesteps 
    with the first $N_1=10^6$ timesteps discarded (see \eqref{structfactN1N2}).
    The middle panel displays its error $\lvert S^{U_x}_{\Delta \mathbf{x},\Delta t}-S^{U_x}_{\Delta \mathbf{x}}\rvert$, where $S^{U_x}_{\Delta \mathbf{x}}$ is the exact result (\ie for $\Delta t\rightarrow 0$).
    The right panel displays the expected error (\ie without statistical errors) computed using the semi-analytic results.}
\end{figure}

We recall that the equilibrium structure factors are defined as
\begin{equation}
    S^{U_x}_{\Delta \mathbf{x},\Delta t} 
    = V \lim_{n\rightarrow\infty}\langle \hat{U}_x(n\Delta t)\hat{U}_x^*(n\Delta t)\rangle,\quad
    S^{U_y}_{\Delta \mathbf{x},\Delta t} 
    = V \lim_{n\rightarrow\infty}\langle \hat{U}_y(n\Delta t)\hat{U}_y^*(n\Delta t)\rangle,
\end{equation}
where $V$ is the volume of the system and $(\hat{U}_x(n\Delta t),\hat{U}_y(n\Delta t))$ is
the Fourier transform of $U^n = (U_x(n\Delta t),U_y(n\Delta t))$.
As mentioned in section~\ref{sec:cov_and_struct_fac}, due to the no-slip boundary conditions, 
we consider the extended domain (see Figure~\ref{figappendix}) to define the Fourier modes.
As a result, for a system with $N\times N$ cells, there are $2N\times 2N$ Fourier modes and
the Fourier spectrum is symmetric with respect to $k_x=0$ and $k_y=0$
(see Figure~\ref{fig:structfact12x12} for the $12\times 12$ case).

Using a simulated time trajectory, the equilibrium structure factors are calculated as follows.
Since the procedure is exactly the same for $S^{U_y}_{\Delta \mathbf{x},\Delta t}$, 
we only explain the case of $S^{U_x}_{\Delta \mathbf{x},\Delta t}$.
When the time trajectory is computed up to $N_2$ timesteps,
we compute the following time average to estimate the ensemble average 
$\langle \hat{U}_x\hat{U}_x^*\rangle$, where the first $N_1$ timesteps are discarded 
to not include non-stationary data:
\begin{equation}
\label{structfactN1N2}
  S^{U_x}_{\Delta \mathbf{x},\Delta t} \approx \frac{V}{N_2 - N_1} \sum_{n= N_1+1}^{N_2}
  \hat{U}_x(n\Delta t)\hat{U}_x^*(n\Delta t).
\end{equation}
Hence, a sufficiently large value of $N_1$ should be used to reduce the systematic error,
whereas the value of $N_2-N_1$ should be large enough to control the level of 
the statistical error.
By the central limit theorem, the magnitude of the statistical error is asymptotically 
proportional to $1/\sqrt{N_2-N_1}$.
The exact structure factor $S^{U_x}_{\Delta \mathbf{x}}$ of the spatially discretized case 
can be computed using $R_{\Delta \mathbf{x}}$ given in \eqref{R_disc}.
Similarly, the theoretical values of the numerical structure factor 
$S^{U_x}_{\Delta \mathbf{x},\Delta t}$,
which one would obtain from \eqref{structfactN1N2} in the limits
$N_1\rightarrow\infty$ and $N_2-N_1\rightarrow\infty$ can be computed 
using the semi-analytic result $R^{(K)}_{\Delta \mathbf{x},\Delta t}$.

We present here the numerical structure factor results of $S^{U_x}_{\Delta \mathbf{x},\Delta t}$ 
that were calculated using the $2$-iteration scheme (\ie $K=2$)
for three different system sizes, $12\times 12$, $16\times 16$, and $32\times 32$ cells.
Parameter values, $\Delta x=\Delta y = 1$, $\beta = \nu\Delta t/\Delta x^2 = 1$, and
$\rho/k_\mathrm{B}T = 1$, were used.
The initial velocity field was set to zero and the trajectory was calculated up to 
$N_2=10^7$ timesteps for the $12\times 12$ and $16\times 16$ cases 
and $N_2=2\times 10^6$ for the $32\times 32$ case. 
To compute the equilibrium structure factor, the first $N_1=10^6$ timesteps were discarded.

Figure~\ref{fig:structfact12x12} shows the numerical structure factor
$S^{U_x}_{\Delta \mathbf{x},\Delta t}$ for the smallest system size.
Since the semi-analytic result of $S^{U_x}_{\Delta \mathbf{x},\Delta t}$ is available
for this case, 
we compare the simulation error
$\lvert S^{U_x}_{\Delta \mathbf{x},\Delta t}-S^{U_x}_{\Delta \mathbf{x}}\rvert$
with the theoretically expected error (\ie without statistical errors) for validation purposes.
We see that the time integration error in the equilibrium structure factor
due to inexact boundary corrections is reasonably small for rather large $\beta = 1$ 
even when one boundary correction is used per timestep.
In fact, the plot of $S^{U_x}_{\Delta \mathbf{x},\Delta t}$ is 
visually indistinguishable from that of the exact result $S^{U_x}_{\Delta \mathbf{x}}$
(\ie for $\Delta t\rightarrow 0$).

It is instructive to observe some features of $S^{U_x}_{\Delta \mathbf{x}}$.
First, for the no-slip boundary and periodic boundary cases, their equilibrium structures
are overall similar but different.
For the periodic boundary case, $S^{U_x}_{\Delta \mathbf{x}}$ is given as 
(see \eqref{S_pbc})
\begin{equation}
    S^{U_x}_{\Delta \mathbf{x}} = \frac{k_\mathrm{B}T}{\rho}\frac{k_y^2}{k_x^2+k_y^2},
\end{equation}
and thus along a ray (\ie $k_y = c k_x$)
the values of $S^{U_x}_{\Delta \mathbf{x}}$ do not change.
However, our no-slip boundary case result shows that the values of $S^{U_x}_{\Delta \mathbf{x}}$
slightly change along a ray for larger values of $k_x$ and $k_y$.
Second, we see that $S^{U_x}_{\Delta \mathbf{x},\Delta t}$ becomes zero for the Fourier modes 
with $k_x=0$.
This is because if $U_x$ is independent of $x$, it must be zero
due to the boundary condition.
In addition, $S^{U_x}_{\Delta \mathbf{x},\Delta t}$ also becomes zero for the Fourier modes 
with $k_y=0$ because these modes are omitted by the projection operator $P$.

\begin{figure}[t!]
\centering
\includegraphics[width=0.65\textwidth]{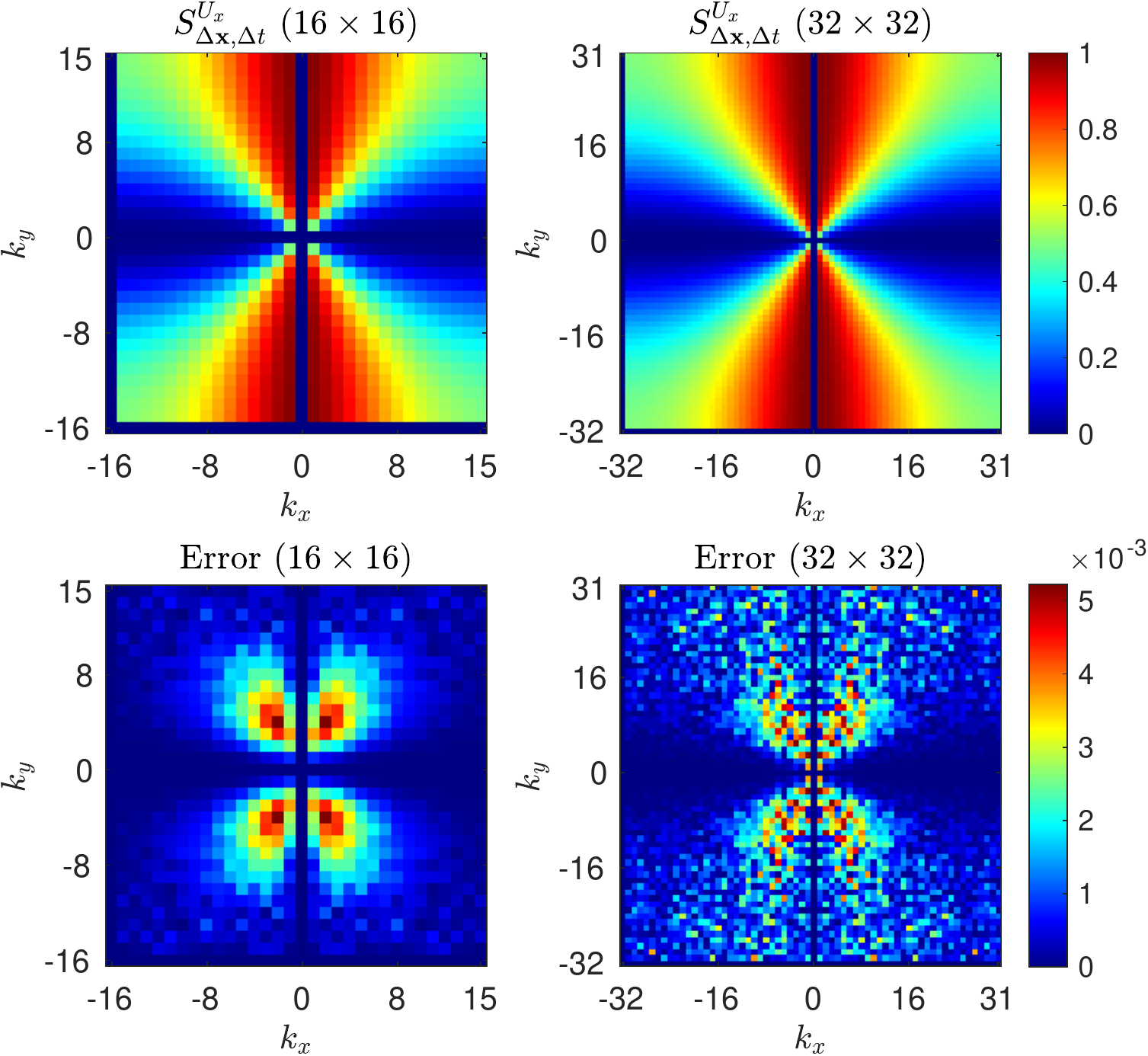}
\caption{\label{fig:structfact16x16v32x32}
    The equilibrium structure factors $S^{U_x}_{\Delta \mathbf{x},\Delta t}$ computed
    for two system sizes, $16\times 16$ (left) and $32\times 32$ (right) cells
    with the same cell sizes $\Delta x = \Delta y = 1$, are shown in the top row.
    The results were obtained using the 2-iteration scheme 
    with $\beta = \nu\Delta t/\Delta x^2 = 1$.
    The corresponding errors 
    $\lvert S^{U_x}_{\Delta \mathbf{x},\Delta t}-S^{U_x}_{\Delta \mathbf{x}}\rvert$
    are shown in the bottom row.
    While the same values of $N_1 = 10^6$ were used for both sizes,
    different values of $N_2$, $10^7$ for $16\times16$ and $2\times 10^6$ for $32\times 32$,
    were used.
    As a result, the level of statistical errors for $32\times 32$ is higher.}
\end{figure}

Figure~\ref{fig:structfact16x16v32x32} shows the results of the larger system sizes.
As in the $12\times 12$ case, the plots of $S^{U_x}_{\Delta \mathbf{x},\Delta t}$ 
are visually indistinguishable from those of $S^{U_x}_{\Delta \mathbf{x}}$.
As the number of cells increases, the plot of $S^{U_x}_{\Delta \mathbf{x}}$ becomes similar to 
that of the continuum structure factor.
The characteristic pattern observed in the error plot of the $12\times 12$ case
appears in the error plots of the larger systems.
Due to the smaller value of $N_2$ for the $32\times 32$ case, the level of statistical errors is 
relatively significant in the error plot.
However, even for this case, the level of statistical errors in the structure factor plot
is completely negligible.


\section{\label{sec_giant_fluct}Simulations of Giant Fluctuations}

In this section, we apply our projection method to simulate the phenomenon of giant fluctuations.
As experimentally observed in space~\cite{GFexp, GFexp2}, 
random advection (due to thermal fluctuations) can induce long-ranged concentration fluctuations when coupled with a concentration gradient in a micro-gravity environment.
Specifically, in the absence of gravity, the nonequilibrium enhancement in the structure factor of the concentration fluctuations
exhibits a power-law divergence, $S^c_\mathrm{neq}(k) \propto k^{-4}$, where $k$ is the wavenumber.
The theoretical explanation of the phenomenon~\cite{OrtizDeZarateSengers2006} is regarded as one of the most significant accomplishments of
the FHD approach.
The incompressible computer simulation of the phenomenon was first performed in~\cite{BalboaUsabiagaBellDelgadoBuscalioniDonvFaiGriffithPeskin2012},
where the velocity equation was solved using the saddle-point system.
In this section, we simulate giant fluctuations with similar settings considered in
\cite{BalboaUsabiagaBellDelgadoBuscalioniDonvFaiGriffithPeskin2012}.
While a careful and systematic investigation on the choice of spatial resolution for given physical parameter values would be in general required in an SPDE simulation study, we use simulation parameter values similar to the ones established in \cite{BalboaUsabiagaBellDelgadoBuscalioniDonvFaiGriffithPeskin2012}.
The goal of this section is to demonstrate that our projection method approach is readily applicable to the velocity equation coupled with the concentration equation
and gives comparable results without the monolithic system.

\subsection{Governing Equations}

\begin{figure}[t]
    \centering 
    \includegraphics[width=0.95\linewidth]{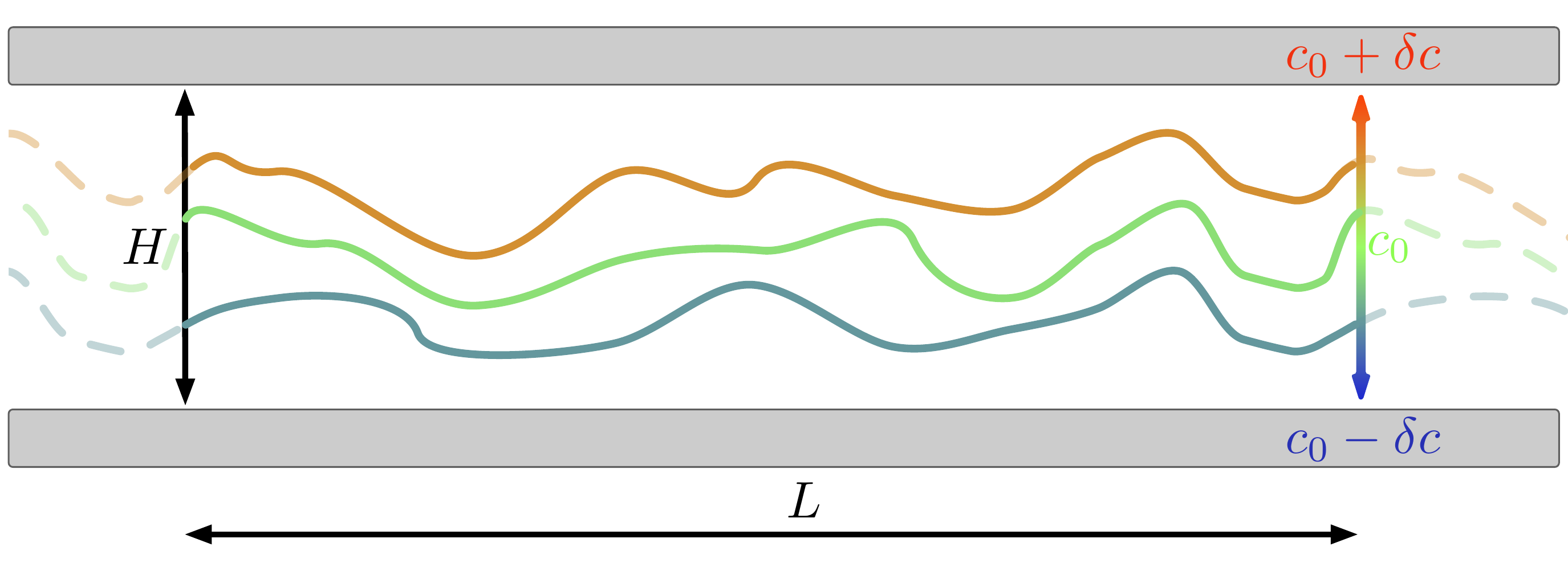}
    \caption{\label{fig:GF_schematic}
    Schematic description of the giant fluctuation phenomenon.
    When a concentration gradient is present in a fluid system in the absence of gravity,
    the concentration field exhibits long-ranged correlations along the directions perpendicular to
    the concentration gradient.
    As a result, diffusing fronts (\ie surfaces of constant concentration) become very rough.
    For a snapshot of diffusing fronts from our giant fluctuation simulation, 
    see Figure~\ref{fig:belle-photo}.}
\end{figure}

Following \cite{BalboaUsabiagaBellDelgadoBuscalioniDonvFaiGriffithPeskin2012}, we consider
a two-dimensional incompressible fluid system confined between two walls in the absence of gravity
(see Figure~\ref{fig:GF_schematic}).
The fluid is a dilute solution, where the concentrations at the walls are held fixed 
with slightly different values. As a result the mean concentration profile has a small concentration gradient.
The governing equations for the velocity $\mathbf{u}$ and the mass fraction $c$ of the solute are 
written as
\begin{equation}
  \label{eq:GFeq}
  \left\{
    \begin{aligned}
     \frac{ \partial \mathbf{u}}{\partial t} &= \nu \Delta \mathbf{u} - \bm{\nabla}\pi + \sqrt{\frac{2k_\mathrm{B}T\nu}{ \rho}} \bm{\nabla} \cdot \bm{\mathcal{W}}_v, \\
      \bm{\nabla} \cdot \mathbf{u} &= 0, \\
      \frac{\partial c}{\partial t} +\mathbf{u} \cdot \bm{\nabla} c &=  \chi \Delta c +
      \sqrt{\frac{2c(1-c)M \chi}{\rho}} \bm{\nabla} \cdot
      \bm{\mathcal{W}}_c,
    \end{aligned}
  \right.
\end{equation}
where $\rho$ is the constant density of the solution, $\nu$ is the kinematic viscosity of the solution, 
$\chi$ is the diffusivity of the solute in the solution, and $M$ is the mass of a solute molecule.
$\bm{\mathcal{W}}_v$ and $\bm{\mathcal{W}}_c$ are independent spatiotemporal Gaussian white noise 
tensor fields.
Note that we assume that viscous effects dominate inertial effects and 
omit the $\mathbf{u}\cdot\bm{\nabla}\mathbf{u}$ term.
The size of the system domain is $L\times H$.
At the walls, Dirichlet boundary conditions are imposed for $c$:
$c=c_0\pm\delta c$ at $y=\pm H/2$, where $c_0$ is the mean mass fraction and $\delta c \ll c_0$.
For the velocity, no-slip boundary conditions are imposed on the walls.
Periodic boundary conditions are imposed for $\mathbf{u}$ and $c$ in the horizontal direction.

\subsection{\label{subsec_GF_lin}Linear Case}

\begin{figure}[t!]
    \centering
    \includegraphics[width=0.48\linewidth]{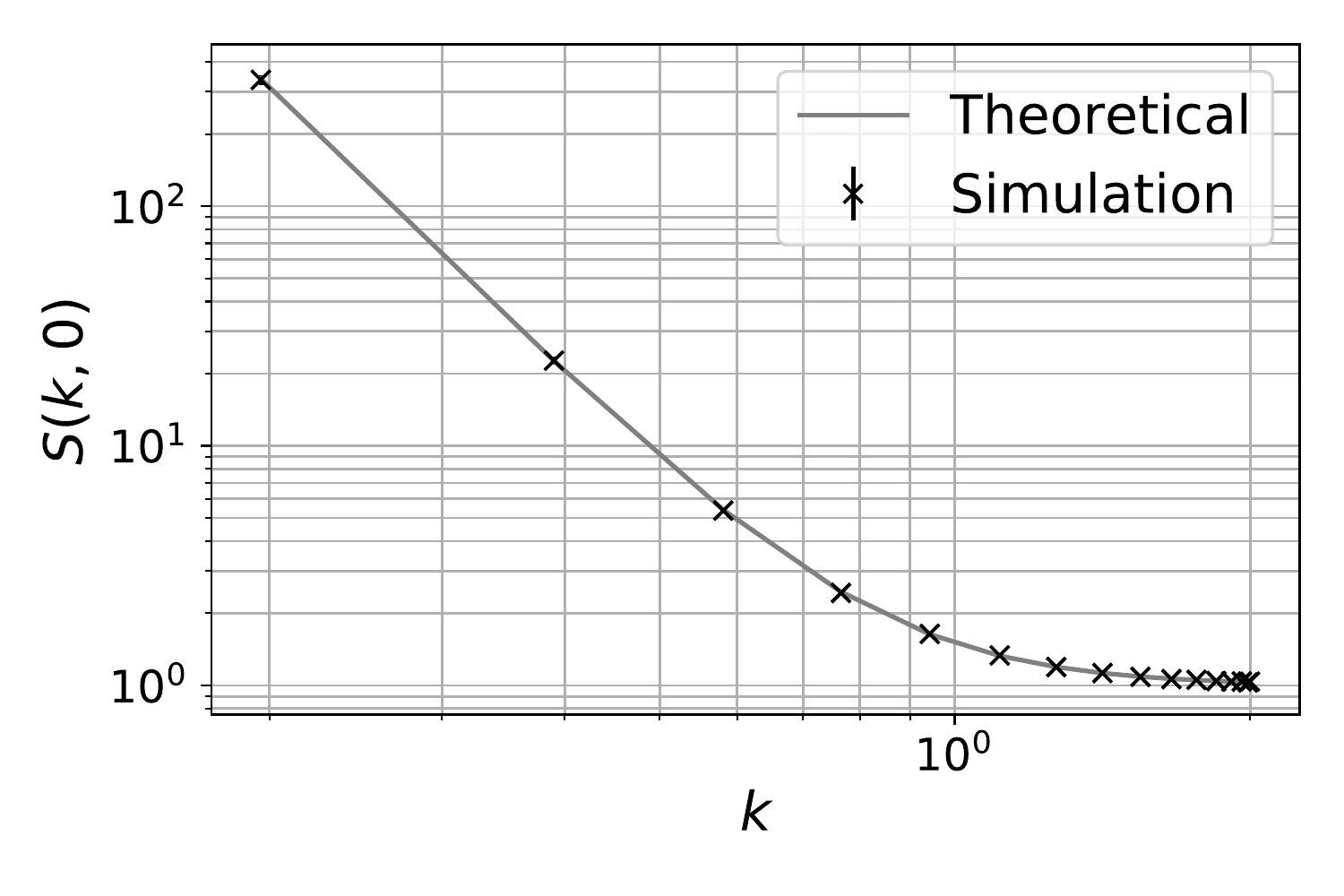}
    \includegraphics[width=0.48\linewidth]{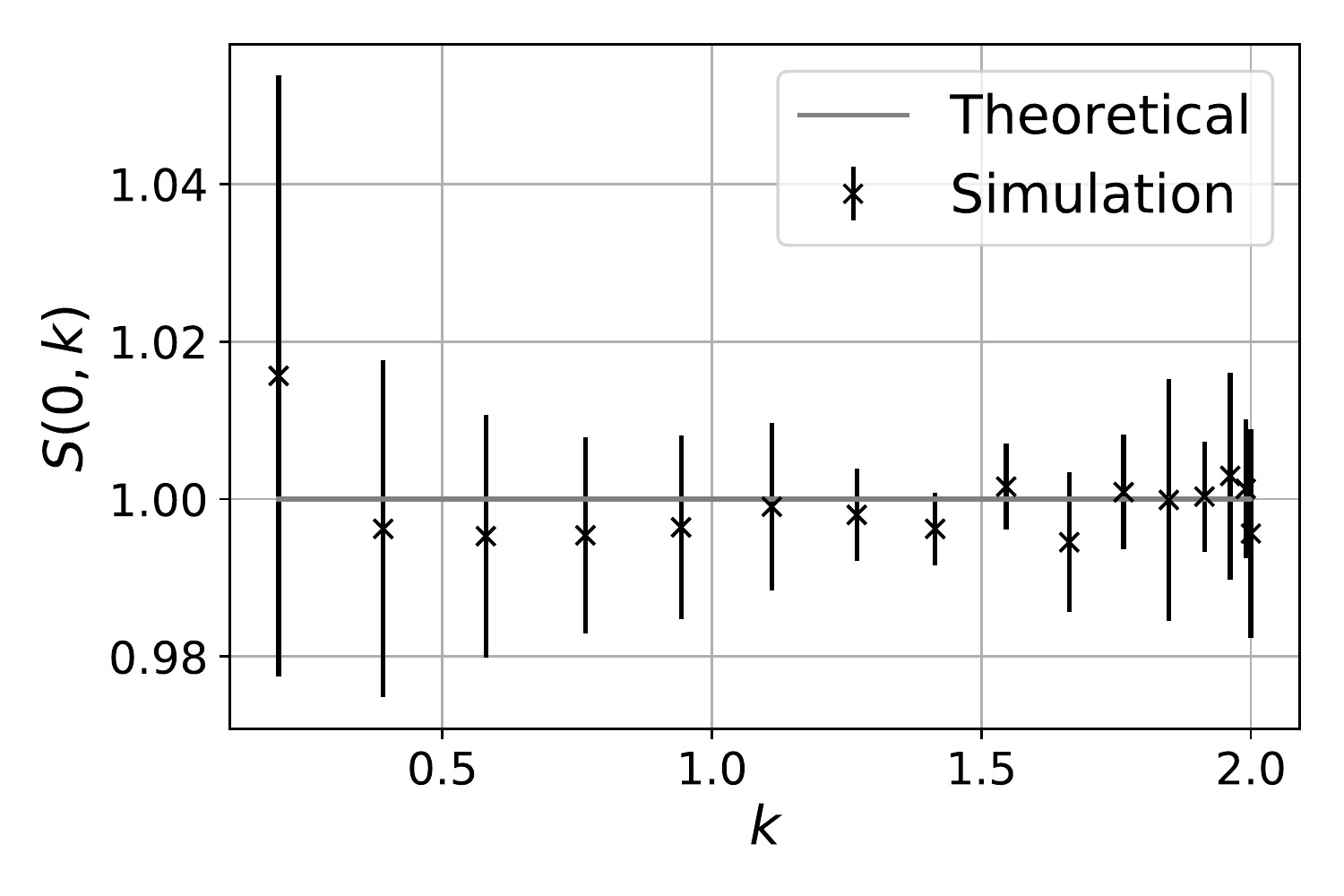}
    \caption{\label{fig:gf_linear}
    Comparison of the simulation results of the concentration structure factor $S^c_\mathrm{simul}$
    with the theoretical results $S^c_{\bm \Delta x}$ for the linear case.
    In the left panel, the power-law divergence is shown for $S^c(k,0)$.
    In the right panel, $S^c(0,k)$ is shown, which corresponds to the equilibrium structure factor
    $S^c_\mathrm{eq}$ (see \eqref{GF_Sc}).
    The linearized periodic system~\eqref{eq:GFeq_linear} is solved for 
    $L = H = 32$, $\Delta x = \Delta y = 1$, $\Delta t=1$, $\mu = \chi = 1$, and $\lvert\overline{\bm{\nabla} c}\rvert = 1$.
    The mean values and error bars are computed from 4 independent runs with $5\times 10^4$ timesteps.
    }
\end{figure}

Due to the Dirichlet boundary conditions and the nonlinear advection term $\mathbf{u}\cdot\bm{\nabla}c$,
analytic results for the structure factor $S^c$ are limited for \eqref{eq:GFeq}~\cite{OrtizdeZaratePelusoSengers2004}.
However, the theoretical prediction of a power-law divergence $S^c_\mathrm{neq}(k) \propto k^{-4}$
can be readily obtained for a linearized periodic system~\cite{OrtizDeZarateSengers2006}.
We first describe this system and summarize the analytic results, 
and then apply our numerical approach to the system and 
compare our numerical results with the analytic results for validation.

We linearize $c$ around the mean concentration profile 
$\bar{c}(y) =c_0 + \left|\overline{{\bm\nabla}c}\right|y$,
where the constant gradient is given as $\overline{{\bm\nabla}c}=(0,2\delta c/H)$.
We denote $c-\bar{c}(y)$ by $\tilde{c}$.
By approximating the advection term and the stochastic term using
constant $\overline{{\bm\nabla}c}$ and $c_0$, respectively, we obtain
\begin{equation}
  \label{eq:GFeq_linear}
  \left\{
    \begin{aligned}
     \frac{ \partial \mathbf{u}}{\partial t} &= \nu \Delta \mathbf{u} - \bm{\nabla}\pi + \sqrt{\frac{2k_\mathrm{B}T\nu}{ \rho}} \bm{\nabla} \cdot \bm{\mathcal{W}}_v, \\
      \bm{\nabla} \cdot \mathbf{u} &= 0, \\
      \frac{\partial \tilde{c}}{\partial t} +\mathbf{u} \cdot \overline{\bm{\nabla} c}  
      &= \chi \Delta \tilde{c} + \sqrt{\frac{2c_0(1-c_0)M \chi}{\rho}} \bm{\nabla} \cdot
      \bm{\mathcal{W}}_c.
    \end{aligned}
  \right.
\end{equation}
Since it is expected that the fluctuational behavior of $\tilde{c}$ does not significantly depend on $y$
under a weak constant concentration gradient, 
we further assume that periodic boundary conditions can be imposed 
for the vertical direction in \eqref{eq:GFeq_linear}.
While the validity of those assumptions needs to be investigated for realistic cases, 
we recall that the linear periodic system exhibits the essential physics of the power-law divergence.
We note that this periodic approximation was suggested and justified in the physics literature on long-range nonequilibrium correlations~\cite{OrtizDeZarateSengers2006}.
The steady-state concentration structure factor is given as~\cite{BalboaUsabiagaBellDelgadoBuscalioniDonvFaiGriffithPeskin2012}
\begin{equation}
  \label{GF_Sc}
  S^c(k,l) = S^c_\mathrm{eq} + S^c_\mathrm{neq}
  = \frac{Mc_0(1-c_0)}{\rho} 
  + \frac{k_\mathrm{B}T \left|\overline{\bm{\nabla} c}\right|^2}{\rho \chi (\chi + \nu)}\frac{k^2}{k^{2} + l^2}\frac{1}{(k^2 + l^2)^{2}}.
\end{equation}
Hence, by setting $l=0$ (\ie considering $\tilde{c}_\perp = H^{-1}\int_{-H/2}^{H/2} \tilde{c}\;dy$),
we see that $S^c_\mathrm{neq} \propto k^{-4}$.
On the other hand, for $k=0$, $S^c$ has only the equilibrium structure factor $S^c_\mathrm{eq}$.

We implemented a numerical scheme to solve \eqref{eq:GFeq_linear}.
Except for the use of our projection method to solve the velocity equation, this scheme strictly follows
the one used in~\cite{BalboaUsabiagaBellDelgadoBuscalioniDonvFaiGriffithPeskin2012}.
That is, for the discretization of the concentration equation, we define the concentration at cell centers
and discretize the Laplacian and stochastic terms in the same manner as for velocity 
but shifted to the regular mesh.
For the discretization of the advection term $\mathbf{u} \cdot \overline{\bm{\nabla} c}$, we interpolate the velocity at cell centers by averaging the neighboring face values.
We employ our projection method to update the velocity field for time discretization and then solve the concentration equation using a semi-implicit Crank--Nicolson scheme.
Since periodic boundary conditions are enforced, no boundary correction for velocity is required (\ie $K=1)$.

For validation, we compare the structure factor results obtained from the numerical simulations
with the theoretical results.
We note that the analytical expression for the corresponding discrete structure factor 
$S^c_{\bm \Delta x}$ is obtained from \eqref{GF_Sc} 
by replacing $k$, $l$, and $\left|\overline{\bm{\nabla} c}\right|^2$
with ${2 \sin (\pi k \Delta x)}/{\Delta x}$,
${2 \sin (\pi l \Delta y)}/{\Delta y}$,
and $\cos(\pi \Delta y) \left|\overline{\bm{\nabla} c}\right|^2$, respectively.
Figure~\ref{fig:gf_linear} shows that the numerical simulation results
match well with the theoretical results.
The left panel clearly shows the power-law divergence of $S^c(k,0)\propto k^{-4}$ for small $k$.
For large $k$, the power-law of the nonequilibrium part $S^c_\mathrm{neq}$ is hidden 
by the equilibrium part $S^c_\mathrm{eq}$ as expected from \eqref{GF_Sc}. 
The right panel shows that $S^c(0,k)$ coincides with the equilibrium structure factor
$S^c_\mathrm{eq}$, which confirms that the fluctuation spectrum parallel to the concentration gradient
is not affected.

\subsection{Realistic Case}

\begin{figure}[t]
\includegraphics[width=0.45\linewidth]{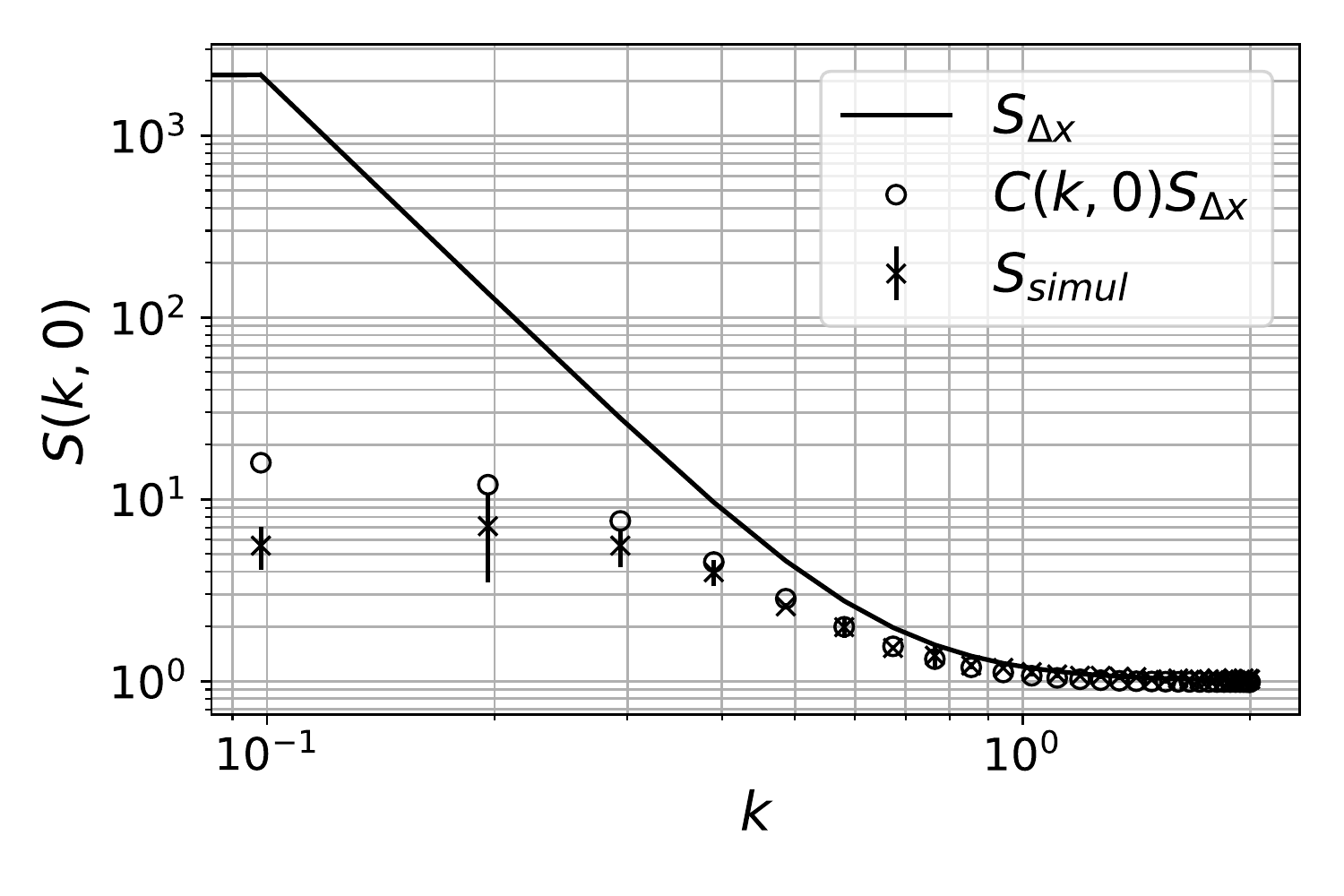}
\includegraphics[width=0.45\linewidth]{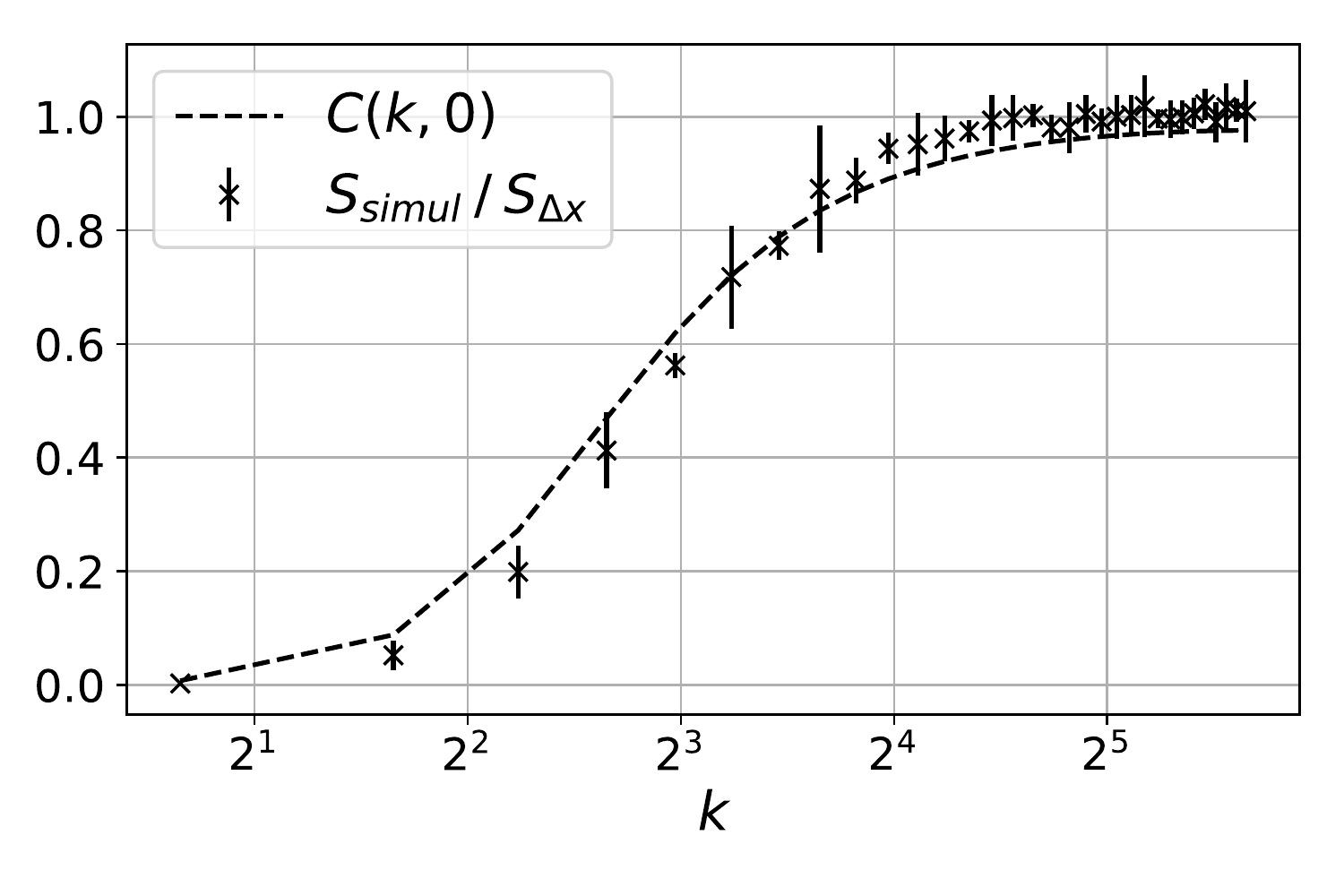}
\caption{\label{figGFreal}
Simulation results of the concentration structure factor $S_\mathrm{simul}$ 
for the giant fluctuation phenomenon.
In the left panel, $S_\mathrm{simul}$ is compared with the theoretical result $S_{\bf\Delta x}$
of the linear case as well as the corrected theoretical result $C(k,0)S_{\bf\Delta x}$.
In the right panel, the approximation for the suppression factor $C(k,0)$ is compared
with $S_\mathrm{simul}/S_{\bf\Delta x}$.
The simulations results were obtained by solving \eqref{eq:GFeq} with
$\Delta x = \Delta y = 1$, $L = 128$, $H = 16$, $\Delta t=0.2$, $\mu=4$, and $\chi = 1$.
The Dirichlet condition was imposed with a gradient of $1$.
The error bars were computed from 4 independent runs with $5000\Delta t$.}
\end{figure}

In this last part, we present numerical results for the original system \eqref{eq:GFeq}.
As for the linear case, our numerical approach follows the one used in
\cite{BalboaUsabiagaBellDelgadoBuscalioniDonvFaiGriffithPeskin2012} 
except that we solve the velocity equation using our projection method.
Because no-slip boundary conditions are enforced in the vertical direction, we must now compute the velocity field using the $K$-iteration scheme ($K>1$). For the discretization of the advection term $\mathbf{u} \cdot \bm{\nabla}c$ and
the implementation of the Dirichlet boundary condition for the concentration, see  
\cite{BalboaUsabiagaBellDelgadoBuscalioniDonvFaiGriffithPeskin2012}.

\begin{figure}[t!]
    \centering \includegraphics[width=\linewidth]{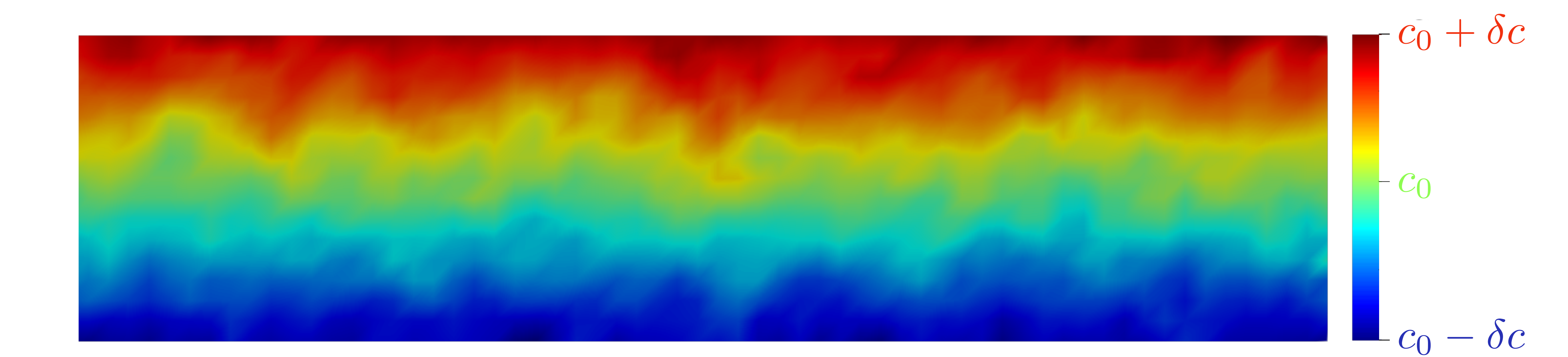}
    \caption{Snapshot of the concentration profile. For the simulation parameters, see the caption of
    Figure~\ref{figGFreal}.}
    \label{fig:belle-photo}
\end{figure}

The concentration boundary conditions are known to suppress the power-law divergence at small $k$.
For the Rayleigh-B\'enard problem for binary fluid mixtures,
the following closed-form approximation for the suppression factor
(\ie the ratio of the realistic structure factor to the linearized one) was obtained~\cite{OrtizdeZaratePelusoSengers2004}:
\begin{align}
  \label{GFsup}
  C(k, l=0) = \frac{(k H)^4}{(k H)^4 + 24.6(k H)^2 + 505.5}.
\end{align}
Although our problem setting is not exactly the same as the Rayleigh-B\'enard problem, 
we test whether similar corrections can be applied. 
Figure~\ref{figGFreal} shows our simulation results for $K=2$.
As expected, we observe that
the Dirichlet boundary condition significantly affects the structure factor at small $k$
and, as a result, the power-law divergence becomes weaker (see the left panel). 
Also, our measurement of the suppression factor agrees well with the above theoretical prediction~\eqref{GFsup} over a wide range of $k$ (see the right panel).
These conclusions are qualitatively the same as the one in
\cite{BalboaUsabiagaBellDelgadoBuscalioniDonvFaiGriffithPeskin2012}.
A snapshot of the concentration profile is shown in Figure~\ref{fig:belle-photo}.

\section{\label{sec_conclusions}Conclusions}

Motivated to incorporate thermal fluctuations into traditional CFD calculations,
we have presented how the projection method on a uniform staggered grid can be adapted 
to include stochastic contributions, and demonstrated that this approach can accurately solve 
the incompressible FHD equations. 
To this end, we have analyzed the resulting equilibrium structure factor of the velocity field, 
or equivalently the steady-state covariance function. 
For periodic boundary conditions, the projection method does not introduce any splitting errors and 
thus it gives exactly the same structure factor as the one that monolithic solvers would give. 
For non-periodic boundary conditions, such as no-slip, splitting errors occur at the boundaries. 
To correct them, we have proposed to use a simple iterative procedure. 
We have shown and verified that the splitting errors converge exponentially with the number of iterations
and the convergence rate depends on the dimensionless number $\beta=\nu\Delta t/\Delta x^2$.

Overall our method succeeds at simulating the incompressible FHD equations
efficiently and accurately without the need to form and solve a monolithic system 
containing the discretization of the momentum and pressure equations. 
The construction of the method demonstrates how other variations of the projection method, and by extension other CFD techniques, can be adapted to incorporate 
thermal fluctuations in fluid simulations at small scales. 
Its application to the simulation of giant fluctuations illustrates how it can be used 
to explore the impact of thermal fluctuations on complex multi-physics fluid flows.
While our numerical analysis based on the equilibrium structure factor and the nonequilibrium simulation study of giant fluctuations confirm that the proposed projection method gives comparable results without the monolithic system, further numerical investigation (\eg based on the dynamic structure factor spectra) would be beneficial to characterize the system behaviors and propagation of numerical errors.

\section*{Acknowledgments}

The authors would like to deeply thank Dr.\ John Bell (Lawrence Berkeley National Laboratory)
for insightful discussions on various aspects of the computational FHD approach
and Matteo Polimeno (UC Merced) for helping with the data visualization. 
C.K.\ also thanks Dr.\ Lei, Yue (UC Merced) for helpful discussions on the mathematical aspect of
Appendix~A.
The authors gratefully acknowledge computing time on the Multi-Environment Computer for Exploration 
and Discovery (MERCED) cluster at UC Merced, which was funded by National Science Foundation Grant
No.~ACI-1429783.

\section*{Credit Author Statement }  

{\bf Marc Mancini}: Investigation, Visualization, Formal Analysis, Validation, Software, Writing - Original Draft. {\bf Maxime Theillard}: Conceptualization, Software, Project administration, Supervision, Writing - Original Draft, Writing - Review $\&$ Editing. {\bf Changho Kim}: Conceptualization, Supervision, Formal Analysis, Methodology, Software, Project administration, Writing - Original Draft, Writing - Review $\&$ Editing.


\appendix

\section{\label{appendix_proj_sdes}Projected Linear SDEs and Steady-State Covariance}

\subsection{Continuous-time case}

In this appendix, we present analytic results on the determination of 
the steady-state covariance matrix $R = \langle \mathbf{x}\mathbf{x}^* \rangle$
for the following form of projected linear SDEs:
\begin{equation}
  \label{projlinSDE}
  \frac{d\mathbf{x}}{dt} = P\left[ A\mathbf{x} + B\boldsymbol{\mathcal{Z}}(t) \right].
\end{equation}
Here, the $n$-dimensional stochastic process $\mathbf{x}(t)$ is driven 
by an $m$-dimensional Gaussian white noise process $\boldsymbol{\mathcal{Z}}(t)$ 
with covariance
$\langle \boldsymbol{\mathcal{Z}}(t) \boldsymbol{\mathcal{Z}}^*(t') \rangle
= C_{\boldsymbol{\mathcal{Z}}} \delta (t-t')$
and its dynamics is projected by an orthogonal projection operator $P$
(\ie $P^2=P$ and $P^*=P$).
Hence, $P$, $A$, $B$, and $C_{\boldsymbol{\mathcal{Z}}}$ are
$n\times n$, $n\times n$, $n\times m$, and $m\times m$ matrices, respectively.
We will make further assumptions on $A$ and initial condition $\mathbf{x}(0)$ below. 

We first summarize standard analytic results for the case of unprojected SDEs:
\begin{equation}
  \label{unprojlinSDE}
  \frac{d\mathbf{x}}{dt} = A\mathbf{x} + B\boldsymbol{\mathcal{Z}}(t).
\end{equation}
We assume that all eigenvalues of $A$ have negative real parts.
The long-time dynamics of $\mathbf{x}(t)$ converge therefore to a steady state 
characterized by a Gaussian distribution
$\rho_\mathrm{ss}(\mathbf{x}) = Z^{-1}\exp(-\frac12\mathbf{x}^*R^{-1}\mathbf{x})$,
where $R = \int \mathbf{x} \mathbf{x}^* \rho_\mathrm{ss}(\mathbf{x}) d\mathbf{x}
= \lim_{t\rightarrow\infty}\langle \mathbf{x}(t) \mathbf{x}^*(t)\rangle$
is the steady-state covariance matrix and $Z$ is the normalization constant.
Since the steady-state process $\mathbf{x}_\mathrm{ss}(t)$ can be expressed as
\begin{equation}
  \mathbf{x}_\mathrm{ss}(t) 
  = \int_{-\infty}^t e^{A(t-s)}B\boldsymbol{\mathcal{Z}}(s) ds,
\end{equation}
it is straightforward to obtain the following expression of its covariance
\begin{equation}
  \label{Cov_intg_unproj}
  R = \langle \mathbf{x}_\mathrm{ss}(0)\mathbf{x}_\mathrm{ss}^*(0) \rangle
  = \int_0^\infty e^{As}B C_{\boldsymbol{\mathcal{Z}}} B^* e^{A^*s} ds.
\end{equation}
However, it is noted that calculating a matrix exponential and
thus evaluating \eqref{Cov_intg_unproj} may not be so straightforward,
particularly if $n$ is large.
An alternative method to compute the covariance matrix $R$ is
to solve a linear system of which $R$ is a solution~\cite{DonevVandenEijndenGarciaBell2010}. 
Since 
\begin{eqnarray}
d R 
= \langle d\mathbf{x} \mathbf{x}^*\rangle + \langle \mathbf{x} d\mathbf{x}^*\rangle 
+ \langle d\mathbf{x} d\mathbf{x}^*\rangle
= A R dt + R A^* dt + B C_{\boldsymbol{\mathcal{Z}}} B^* dt = 0,
\end{eqnarray}
$R$ satisfies
\begin{equation}
\label{Cov_rel_unproj}
A R + R A^* = -B C_{\boldsymbol{\mathcal{Z}}} B^*.
\end{equation}
This linear system has the form of the Lyapunov equation and 
its solubility and uniqueness is guaranteed when the eigenvalues of $A$ have 
negative real parts~\cite{Simoncini2016}.
It is noted that a typical case is that $A$ is negative definite.

We now consider the projected SDEs~\eqref{projlinSDE} with the assumption that
$A$ is negative definite.
As in the unprojected case, the following linear system can be easily obtained:
\begin{equation}
  \label{Cov_rel_proj}
  P A R + R A^* P = -P B C_{\boldsymbol{\mathcal{Z}}} B^* P.
\end{equation}
However, except for the trivial case where $P$ is the identity matrix $I$,
some eigenvalues of the matrix $P A$ are zero and the uniqueness of solutions
of \eqref{Cov_rel_proj} is no longer guaranteed.
The reason that \eqref{Cov_rel_proj} has many solutions can be explained as follows.
For the initial condition $\mathbf{x}(0) = P \mathbf{x}(0) + (I-P) \mathbf{x}(0)$,
since \eqref{projlinSDE} only updates the projected portion $P\mathbf{x}(0)$,
the steady-state covariance $R = \langle \mathbf{x}\mathbf{x}^* \rangle$ 
depends on $(I-P) \mathbf{x}(0)$.

By assuming that the initial state $\mathbf{x}(0)$ belongs to the projected space,
\ie $P\mathbf{x}(0)=\mathbf{x}(0)$, 
we can restrict the dynamics of $\mathbf{x}(t)$ within the projected space.
In this case, $R$ must satisfy an additional condition
\begin{equation}
    \label{Cov_rel_proj2}
    PR = RP = P,
\end{equation}
and the two conditions \eqref{Cov_rel_proj} and \eqref{Cov_rel_proj2} guarantee
the uniqueness of $R$.
This can be explained by the fact that $PA$ becomes negative definite 
in the projected space: for $\mathbf{x}=P\mathbf{x}$,
\begin{equation}
    \mathbf{x}^* PA \mathbf{x}
    = (P\mathbf{x})^* PA (P\mathbf{x})
    = (P\mathbf{x})^* A (P\mathbf{x})<0.
\end{equation}
This result is useful when a candidate expression for the steady-state covariance can
be suggested (\eg by physical intuition).
By showing that it satisfies both \eqref{Cov_rel_proj} and \eqref{Cov_rel_proj2},
one can guarantee that it is indeed the steady-state covariance.

\subsection{Discretized case}

We can also obtain a similar result for the following linear stochastic recurrence
relations:
\begin{equation}
  \label{Cov_rel_proj_recurr}
  \mathbf{x}^{n+1} = P\left[A \mathbf{x}^n + B \boldsymbol{\mathcal{Z}}^n\right].
\end{equation}
Here, all assumptions remain the same as above except that
$\langle\boldsymbol{\mathcal{Z}}^n \boldsymbol{\mathcal{Z}}^{n'}\rangle 
= C_{\boldsymbol{\mathcal{Z}}} \delta_{n,n'}$
and $R = \lim_{n\rightarrow\infty} \langle \mathbf{x}^n(\mathbf{x}^n)^*\rangle$.
From the stationarity of $R$, the following linear system is obtained~\cite{DonevVandenEijndenGarciaBell2010}:
\begin{equation}
  \label{Cov_rel_proj_disc}
  PA R (PA)^* - R = -PB C_{\boldsymbol{\mathcal{Z}}}(PB)^*.
\end{equation}
As in the continuous case, this system has many solutions due to the null space of $P$.
When $P\mathbf{x}^0 = \mathbf{x}^0$ is assumed, the resulting $R$ satisfies
\eqref{Cov_rel_proj2}.
For a negative definite $A$, it can be shown that \eqref{Cov_rel_proj_disc} and
\eqref{Cov_rel_proj2} uniquely determine $R$.

\subsection{Reduced linear systems}

We have shown that the steady-state covariance
$R=\lim_{t\rightarrow\infty}\langle\mathbf{x}(t)\mathbf{x}^*(t)\rangle$ 
is the unique solution of \eqref{Cov_rel_proj} and \eqref{Cov_rel_proj2} 
if the time evolution of $\mathbf{x}(t)$ is given by \eqref{projlinSDE}
with the condition $P\mathbf{x}(0)=\mathbf{x}(0)$.
In this case, by restricting the range of $\mathbf{x}(t)$ to the projected space,
we can obtain a reduced linear system for $R$.

This can be done by defining the transformation $\tilde{\mathbf{x}}=V^*\mathbf{x}$
with $V=[\mathbf{v}_1,\dots,\mathbf{v}_r]$.
Here, $r$ is the rank of $P$ and $\{\mathbf{v}_i\}_{i=1,\dots,r}$ are 
a set of $r$ orthonormal eigenvectors of $P$ whose eigenvalues are one.
Using $P=VV^*$ and $I=V^*V$, it is easy to show that
$\tilde{R}=
\lim_{t\rightarrow\infty}\langle\tilde{\mathbf{x}}(t)\tilde{\mathbf{x}}^*(t)\rangle = V^*RV$
satisfies
\begin{equation}
\label{Cov_rel_proj_reduced}
\tilde{A}\tilde{R} + \tilde{R}\tilde{A}^*
= - V^* B C_{\boldsymbol{\mathcal{Z}}} B^* V,
\end{equation}
where $\tilde{A}=V^*AV$.
By vectorizing matrices, \eqref{Cov_rel_proj_reduced} can be expressed as
\begin{equation}
\label{big_sys}
\left(I\otimes\tilde{A}+\tilde{A}\otimes I\right)\mathrm{vec}(\tilde{R})
=-\mathrm{vec}(V^* B C_{\boldsymbol{\mathcal{Z}}} B^* V),
\end{equation}
where $\otimes$ denotes the Kronecker product.
By inverting this $r^2\times r^2$ linear system, $\tilde{R}$ is obtained and 
thus $R=V\tilde{R}V^*$ can be computed.

The same procedure applies to the discretized case.
Assuming \eqref{Cov_rel_proj2}, \eqref{Cov_rel_proj_disc} can be written as
\begin{equation}
    \tilde{A}\tilde{R}\tilde{A}^*-\tilde{R} = - V^*BC_{\boldsymbol{\mathcal{Z}}}B^*V,
\end{equation}
and its vectorized equation is given as
\begin{equation}
    \label{biglinsysdisc}
    \left(\tilde{A}\otimes\tilde{A}-I\right)\mathrm{vec}(\tilde{R})
    = -\mathrm{vec}(V^* B C_{\boldsymbol{\mathcal{Z}}} B^* V).
\end{equation}
Hence, $R=V\tilde{R}V^*$ can be computed from the solution $\tilde{R}$.


\section{\label{appendix_proof_comm}Proof of Commuting Relation (\ref{commutLP})}

\begin{figure}[t!]
    \centering
    \includegraphics[width=0.6\linewidth]{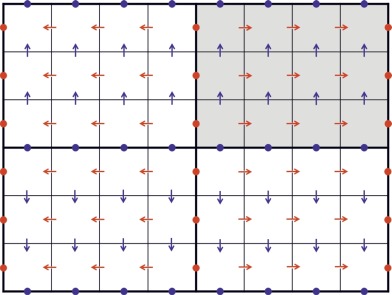}
    \caption{\label{figappendix}
    The velocity field in the extended domain.
    The shaded region depicts the original domain with $4\times 3$ cells.
    Red and blue circles indicate that zero velocities are assigned to $U_x$ and $U_y$, respectively,
    on the boundaries. 
    Red and blue arrows are located at faces where $U_x$ and $U_y$, respectively, are defined 
    inside the domain.
    The directions of the arrows show how the velocities in the original domain are mirrored.}
\end{figure}

By showing that there exists a common eigenbasis, one can show that two operators commute. 
For example, in the case of the periodic boundary case, such a common eigenbasis for
the Laplacian operator and the projection operator can be easily constructed 
from the Fourier modes $(u e^{i(kx+ly)}, v e^{i(kx+ly)})$.
For our case, a similar procedure can be used.
However, due to a different boundary situation, sine-cosine modes
$(u \sin kx \cos ly, v \cos kx \sin ly)$ are used instead to construct a common eigenbasis.

We first show that these sine-cosine modes constitute a complete eigenbasis
of the operator $L_0+B$ as follows.
By the definition of $B$, see \eqref{LL0B}, computing $(L_0+B)U$ is equivalent to 
computing the vector Laplacian of $U$ under the assumption that
its prescribed tangential velocity components on the boundaries
(empty squares in Figure~\ref{fig:BC}) have the same values as the velocity components 
at the face half a grid spacing inward (filled triangles closest to those empty squares).
This implies that the corresponding velocity components in ghost cells (empty triangles)
also have the same values.
Thus, by considering mirror images of $U$ with respect to each boundary (see Figure~\ref{figappendix}), 
we can construct an extended system where periodic boundary conditions are satisfied.
Then it is easy to see that the sine-cosine modes with the form given above form
a complete set of eigenfunctions of $L_0+B$ due to the mirror symmetry and 
the zero normal velocity condition.

We then observe that the action of $P$ does not introduce other modes.
More specifically, we have
$P(u \sin kx \cos ly, v \cos kx \sin ly) = (w \sin kx \cos ly, z \cos kx \sin ly)$,
where
\begin{equation}
  \label{paction}
  \begin{pmatrix} w \\ z \end{pmatrix}
  =\frac{1}{k^2+l^2}
  \begin{pmatrix} l^2 & -k l \\ -k l & k^2 \end{pmatrix}
  \begin{pmatrix} u \\ v \end{pmatrix}.
\end{equation}
Hence, for each mode, we can find simultaneous eigenfunctions of $L_0+B$ and $P$
by diagonalizing the matrix in \eqref{paction}.
It is interesting to see that the matrix is identical to the one for the periodic boundary case.

\bibliographystyle{abbrv} \bibliography{manuscript.bbl}

\end{document}